\documentclass[showpacs,prc,twocolumn,floatfix,aps]{revtex4-1}
\usepackage[dvips]{color, graphicx}
\usepackage{epstopdf}
\usepackage{amssymb} 
\usepackage{amsmath} 
\usepackage{braket}
\usepackage{bm}
\usepackage{multirow}

\graphicspath{{CDBonn_Figures/}}

\begin{document}
\title{Microscopic self-energy of ${}^{40}$Ca from the charge-dependent Bonn potential}
\date{\today}
\author{H. Dussan$^{1}$}
\author{S. J. Waldecker$^{1}$}
\author{W. H. Dickhoff$^{1}$}
\author{H. M\"{u}ther$^{2}$}
\author{A. Polls$^{3}$}
\affiliation{Department of Physics$^{1}$, Washington University,
St.~Louis, Missouri 63130, USA}
\affiliation{Institut f\"{u}r Theoretische Physik$^{2}$, Universit\"{a}t T\"{u}bingen, D-72076, Germany}
\affiliation{ Departament d'Estructura i Constituents de la Mat\`{e}ria and Institut de Ci\`{e}nces del Cosmos$^{3}$, Universitat de Barcelona, Avda. Diagonal 647, E-8028 Barcelona, Spain}

\begin{abstract}
The effects of short-range correlations on the nucleon self-energy in $^{40}$Ca are investigated using the charge-dependent Bonn (CDBonn) interaction. Comparisons are made with recent results for the self-energy of $^{40}$Ca derived from the dispersive optical-model (DOM). 
Particular emphasis is placed on the non-locality of the imaginary part of the microscopic self-energy which suggests that future DOM analyses should include this feature.
In particular, data below the Fermi energy appear sensitive to the implied orbital angular momentum dependence of the self-energy.
Quasiparticle properties obtained for the CDBonn interaction are substantially more mean-field-like than the corresponding DOM results with spectroscopic factors larger by about 0.2 \textit{e.g.}
Reaction cross sections obtained from the microscopic self-energy for scattering energies up to 100 MeV indicate that an adequate description of volume absorption is obtained while a considerable fraction of surface absorption is missing. 
The analysis of the non-locality of the imaginary part of the microscopic self-energy suggests that a simple gaussian provides an adequate description, albeit with rather large values for $\beta$, the non-locality parameter.
\end{abstract}

\maketitle

\section{Introduction}
Calculations of single-particle (sp) observables in nuclei starting from realistic nucleon-nucleon (NN) interactions have demonstrated that several energy scales influence these quantities~\cite{Dickhoff04}. 
Both short-range as well as tensor physics associated with energy scales far above and below the Fermi energy, in addition to long-range physics associated with the nuclear surface, and most influential near the Fermi energy, determine single-nucleon motion in the nucleus. 
This interpretation is well established for nuclei near stability where $(e,e'p)$ experiments have yielded unambiguous results approaching absolute spectroscopic factors~\cite{Lapikas93}.
The observed reduction of the sp strength near the Femi energy of about 35\% can only be quantitatively explained by a combination of the effects of long-range and short-range correlations (SRC)~\cite{Dickhoff04,Barb09}.
While it appears that the bulk of the depletion is associated with the coupling of sp motion to low-lying collective states and giant resonances that are mostly associated with surface phenomena, there is an approximately modest 10\% depletion due to the short-range and tensor force of the NN interaction calculated from the microscopic self-energy obtained for ${}^{16}$O~\cite{Muther94,Muther95,Polls95}.
These calculations also demonstrate an accompanying modest appearance of high-momentum components in the ground state.
The presence of about 10\% high-momentum components does have a significant influence on the binding energy of the ground state, in some cases accounting for about 65\% of the binding energy~\cite{Muther95}.

While the details of these high-momentum components depend sensitively on the chosen NN interaction, experimental evidence confirms their presence in about the amount predicted~\cite{Rohe04} although the details of their energy dependence require further study~\cite{Barbieri04b}.
Calculations for ${}^{16}$O employing other methods that include the effect of SRC confirm that they remove strength from valence orbits by about 10\%~\cite{radici94,fabrocini01}.
The presence of high-momentum components in the ground state for nuclei heavier than ${}^{16}$O is usually obtained by performing a local-density approximation of the nuclear-matter spectral function obtained by the correlated-basis-function method at several densities~\cite{sick94,vanneck95}.
It appears that only the Green's function calculations of Refs.~\cite{Muther94,Muther95,Polls95} can generate spectral functions directly for such nuclei.

Green's function calculations generate self-energies that can also be employed at positive energy where they can be transformed into the so-called reducible self-energy that is equivalent with the elastic-nucleon-scattering $\mathcal{T}$-matrix~\cite{Bell59}. 
Elastic-nucleon-scattering data are usually analyzed with so-called optical potentials that provide quite accurate representations of the data~\cite{Becchetti69,Varner91,Koning03}.
Optical potentials are normally formulated in terms of local potentials that have no immediate interpretation as nucleon self-energies. 
A close connection to the nucleon self-energy has been developed in the dispersive-optical-model (DOM) proposed by Mahaux and Sartor~\cite{Mahaux86,Mahaux91}.
By employing the dispersion relation between the real and imaginary part of the self-energy in a subtracted form, only the imaginary part of the self-energy and the real part at the subtraction point (typically the Fermi energy) are required to provide a fit to the available data.
These include not only elastic-scattering data but also bound-state information, thereby providing a natural link between nuclear reactions and nuclear structure when appropriate assumptions are made about the relation of the imaginary part above and below the Fermi energy~\cite{Mahaux91}.
Standard assumptions about the functional form of these potentials include surface and volume contributions with conventional form factors.
The parameters of these potentials are then constrained by a fit to the available data.
Recent implementations of this approach attempt a simultaneous fit to nuclei with different nucleon asymmetry thereby providing access to extrapolations to more exotic nuclei~\cite{Charity06,Charity07,Charity11}.
An extension of the DOM was recently introduced which enhances the domain of its applicability for data below the Fermi energy~\cite{Dickhoff10a}.
Conventional DOM analyses transform the non-local real part of the self-energy at the Fermi energy into a local but energy-dependent potential according to a standard procedure~\cite{Perey62}.
Such an energy-dependence distorts the normalization of solutions of the Dyson equation which is of particular relevance for the calculation of the spectral function below the Fermi energy.
By returning to a non-local form and renormalizing the imaginary part in a well-defined manner~\cite{Mahaux91}, it is possible to obtain properly normalized results for spectral functions, the one-body density matrix, \textit{etc.}~\cite{Dickhoff10a}.

Since all implementations of the DOM have assumed a local representation of the imaginary part of the self-energy, it is of particular relevance to investigate the non-local content of microscopically calculated self-energies that are based on a realistic interaction.
A recent investigation of this type focused on the coupling of the nucleon at low energy~\cite{seth11}. 
Such long-range correlations are well-described by the Faddeev-random-phase approximation (FRPA)~\cite{Barb:1,Barb:2,Barbieri07,Barbieri09a,Barb09,Degroote11}.
The analysis of Ref.~\cite{seth11} suggests that at low energy, mostly representing the coupling to surface excitations, there is ample evidence of a substantial non-local contribution to the nucleon self-energy.
In the present work we extend this investigation to study the predominantly volume coupling associated with the contribution of SRC as determined by the \textit{ab initio} strategy of Refs.~\cite{Muther94,Muther95,Polls95}.

We employ the CDBonn interaction~\cite{cdbonn,cdbonna} in the present work which is a relatively soft interaction~\cite{Rios09}.
At the same time, we establish how such an \textit{ab initio} optical potential fares in the description of elastic scattering.
Since most recent  DOM results include ${}^{40}$Ca, we have also performed the present calculations for this nucleus.
We note that fully microscopic calculations of the optical potential for finite nuclei are sorely lacking and most applications involve a detour through nuclear matter accompanied by a suitable local-density approximation~\cite{Jeukenne77}.
The work reported in Ref.~\cite{Dickhoff10a} demonstrates that the DOM is capable of incorporating SRC in its description of the nucleon self-energy. 
Its current implementation however, does not yet properly account for the effect that SRC have on the distribution of sp strength, in particular that high-momenta appear increasingly likely at more negative energies~\cite{Muther94,Muther95,Polls95}.
We therefore also consider the high-momentum components generated by the CDBonn interaction in ${}^{40}$Ca to clarify this issue.

In Sec.~\ref{sec:theory} we introduce the essential steps involved in the microscopic calculation of the self-energy for ${}^{40}$Ca.
While the initial step involves a nuclear-matter calculation, the subsequent steps correctly incorporate the propagation of nucleons in the finite system under consideration.
The relevant quantities that can be derived by solving the Dyson equation above and below the Fermi energy are also summarized there.
Results for spectral functions, momentum distributions, natural orbits, charge density, energy sum rule, and elastic scattering are presented in Sec.~\ref{sec:results}.
Whenever relevant, these results are compared with corresponding DOM quantities.
We also attempt to identify properties of the non-local imaginary part of the microscopic self-energy for future implementation of DOM analyses.
Conclusions are presented in Sec.~\ref{sec:conc}.

\section{Theory}
\label{sec:theory}
\subsection{Effective interaction}
\label{sec:gmat}
The microscopic calculation of the nucleon self-energy proceeds in two steps, as employed in Refs.~\cite{Muther94,Muther95,Polls95}.
We start with an outline of the general idea and initially employ a schematic formulation.
A diagrammatic treatment of SRC always involves the summation of ladder diagrams.
When only particle-particle (pp) intermediate states are included, the resulting effective interaction is the so-called $\mathcal{G}$-matrix.
The corresponding calculation for a finite nucleus (FN) can be represented in operator form by
\begin{equation}
\mathcal{G}_{FN}(E)= V + V G^{pp}_{FN}(E) \mathcal{G}_{FN}(E) ,
\label{eq:gmatfn}
\end{equation}
where the noninteracting propagator $G^{pp}_{FN}(E)$ represents two particles above the Fermi sea of  the finite nucleus taking into account the Pauli principle.
The simplest implementation of $G^{pp}_{FN}$ involves plane-wave intermediate states (possibly orthogonalized to the bound states).
Even such a simple assumption leads to a prohibitive calculation to solve Eq.~(\ref{eq:gmatfn}) and subsequently generate the relevant real and imaginary part of the self-energy over a wide range of energies above and below the Fermi energy.
We are not aware of any attempt at such a direct solution at this time, except for the use of the $\mathcal{G}$-matrix as an effective interaction at negative energy.
Instead, we employ a strategy developed in Refs.~\cite{Bonatsos89,Borromeo92} that first calculates a $\mathcal{G}$-matrix in nuclear matter at a fixed density and fixed energy according to
\begin{equation}
\mathcal{G}_{NM}(E_{NM})= V + V G^{pp}_{NM}(E_{NM}) \mathcal{G}_{NM}(E_{NM}) .
\label{eq:gmatnm}
\end{equation}
The energy $E_{NM}$ is chosen below twice the Fermi energy of nuclear matter for a kinetic energy sp spectrum and the resulting $\mathcal{G}_{NM}$ is therefore real.
Formally solving Eq.~(\ref{eq:gmatfn}) in terms of $\mathcal{G}_{NM}$ can be accomplished by 
\begin{eqnarray}
\mathcal{G}_{FN}(E) &=& \mathcal{G}_{NM} 
\label{eq:gmatfnp} \\
&+&   \mathcal{G}_{NM} \left\{G^{pp}_{FN}(E) - G^{pp}_{NM} \right\} \mathcal{G}_{FN}(E) ,
\nonumber
\end{eqnarray}
where the explicit reference to $E_{NM}$ is dropped.
The main assumption to make the self-energy calculation manageable is to drop all terms higher than second order in $\mathcal{G}_{NM}$, leading to
\begin{eqnarray}
\mathcal{G}_{FN}(E) &=& \mathcal{G}_{NM} - \mathcal{G}_{NM}  G^{pp}_{NM}  \mathcal{G}_{NM}
\nonumber \\
&+&   \mathcal{G}_{NM} G^{pp}_{FN}(E) \mathcal{G}_{NM} ,
\label{eq:gmatfnq}
\end{eqnarray}
where the first two terms are energy-independent.
Since a nuclear-matter calculation already incorporates all the important effects associated with SRC, it is reasonable to assume that the lowest-order iteration of the difference propagator in Eq.~(\ref{eq:gmatfnq}) represents an accurate approximation to the full result.
This assertion does require further confirmation in future studies.

The Bethe-Goldstone equation for $\mathcal{G}_{NM}$ reads in the appropriate basis
\begin{eqnarray}
\lefteqn{
\left\langle k \ell \right| \mathcal{G}_{NM}^{S J_S K L T} \left| k' \ell' \right\rangle
 =
  \left\langle k \ell   \right| V^{S J_S K L T} \left| k' \ell' \right\rangle  
  } \hspace{.5in}   & &
\label{eq:betheg}\\
 & &  + \frac{1}{2} \sum_{\ell''} \int dk'' (k'')^2 \; 
             \left\langle k \ell \right| V^{S J_S K L T} 
\left| k'' \ell'' \right\rangle
\nonumber\\  & & ~
\frac{\bar{Q} (K,k'')}{E_{NM}- \frac{\hbar^2 K^2}{4m} - \frac{\hbar^2 k''^2} 
{2m}} 
  \left\langle k'' \ell'' \right| \mathcal{G}_{NM}^{S J_S K L T} \left| k' \ell' \right\rangle
   \; .
\nonumber
\end{eqnarray}
The variables $k$, $k'$, and $k''$ denote the relative wave vectors 
between the two nucleons,
$\ell$, $\ell'$, and $\ell''$ the orbital angular momenta 
for the relative motion,
$K$ and $L$ the corresponding quantum numbers for the center-of-mass
motion, $S$ and $T$ the total spin and isospin, and $J_S$ is obtained
by coupling the orbital angular momentum of the relative motion to the spin
$S$.
We note that $\bar{Q}(K,k)$ is the angle-averaged form of the product of
step-functions that allows a partial wave decomposition.
The implied sole dependence of the $G$-matrix on the magnitude of $\bm{K}$,
the center-of-mass wave vector, also ensures that the solution of
Eq.~(\ref{eq:betheg}) does not depend on $L$.
The $L$-label is kept however, to facilitate the recoupling to individual
orbital angular momentum states, as discussed below.
Equation~(\ref{eq:betheg}) generates an appropriate solution
of two-body short-range dynamics but the resulting matrix elements
require further manipulation before becoming useful for the finite nucleus.

\begin{figure}
\includegraphics[width=3.2in]{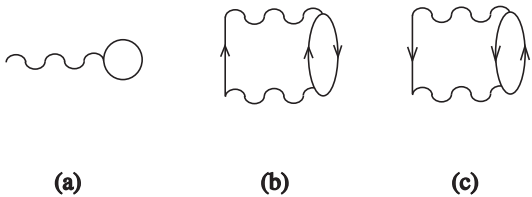}
\caption{Graphical representation of the BHF $(a)$, the 
two-particle--one-hole contribution (b) and one-particle--two-hole term (c) to the
self-energy of the nucleon. The $\mathcal{G}_{NM}$-matrix is indicated by the wiggly line.}
\label{fig:diag}
\end{figure}

The self-energy contribution of the lowest-order term $\mathcal{G}_{NM}$ in Eq.~(\ref{eq:gmatfnq}) is shown in Fig.~\ref{fig:diag}$(a)$ and is similar to a Brueckner-Hartree-Fock (BHF) self-energy. 
While strictly speaking the genuine 
BHF approach involves self-consistent sp wave functions, as in the HF 
approximation, the
main features associated with using the $\mathcal{G}_{NM}$-matrix of Eq.~(\ref{eq:betheg})
are approximately the same when employing a summation over the occupied harmonic oscillator states of ${}^{40}$Ca. Hence we will use the BHF abbreviation.
The correction term involving the second-order in $\mathcal{G}_{NM}$ calculated in nuclear matter is also static and can be obtained from the second term in Eq.~(\ref{eq:betheg}) by replacing the bare interaction by $\mathcal{G}_{NM}$. The corresponding self-energy is also real and generated by summing over the occupied oscillator states in the same way as for the BHF term.

The second-order term containing the correct energy-dependence for $\mathcal{G}_{FN}$ in Eq.~(\ref{eq:gmatfnq}) can now be used to construct the self-energy contribution depicted in Fig.~\ref{fig:diag}(b), representing the coupling to two-particle--one-hole (2p1h) states.
In the calculation harmonic oscillator states for the occupied (hole) 
states and plane waves for the intermediate unbound particle 
states are assumed, incorporating the correct energy and density dependence 
characteristic of a finite nucleus $\mathcal{G}_{FN}$-matrix.
In a similar way, one can construct the second-order self-energy contribution depicted in Fig.~\ref{fig:diag}(c) which has an imaginary part below the Fermi energy and includes the coupling to one-particle--two-hole (1p2h) states.

Calculations of this kind require several basis transformations, including the one from relative and center-of-mass momenta with corresponding orbital angular momenta to two-particle states with individual momenta and orbital angular momentum. 
Complete details can be found in Refs.~\cite{Borromeo92,Muther95}.
In practice the imaginary parts associated with diagrams (b) and (c) of Fig.~\ref{fig:diag} are employed to obtain the corresponding real parts by employing the appropriate dispersion relation.
The resulting (irreducible) self-energy then reads
\begin{eqnarray}
\Sigma^* & = & \Sigma^*_{BHF} + \Delta\Sigma^* \label{eq:defsel}  \\
& = & \Sigma^*_{BHF} + 
\left( \textrm{Re}~\Sigma^*_{2p1h} - \Sigma^*_{c} + \textrm{Re}~\Sigma^*_{1p2h} 
\right) \nonumber \\
&+& i \left( \textrm{Im}~\Sigma^*_{2p1h} + \textrm{Im}~\Sigma^*_{1p2h} \right) \; 
 \nonumber
\end{eqnarray}
in obvious notation.
This self-energy is employed in the sp basis denoted by states $\ket{\left\{ k (\ell \frac{1}{2})  j m_j\right\}}$, characterized by wave vector, orbital, spin, total angular momentum and its projection (suppressing isospin).
We note that the quantum numbers $\ell,j$ and $m_j$ are conserved and the self-energy does not depend on $m_j$. 
We will make use of isospin conservation for ${}^{40}$Ca but include the Coulomb contribution when discussing results for protons.

\subsection{Solution of the Dyson equation}
\label{sec:Dyson}
The sp propagator in momentum space can be obtained from the following version of the Dyson equation~\cite{Dickhoff08}
\begin{eqnarray}
G_{\ell j}(k, k^{\prime}; E) &=& \frac{\delta( k - k^{\prime})}{k^2}G^{(0)}(k; E)  
 \label{eq:gdys1}  \\
		&+&		      G^{(0)}(k; E)\Sigma_{\ell j}(k, k^{\prime}; E)G^{(0)}(k; E) ,
	\nonumber	
\end{eqnarray}
where $G^{(0)}(k; E ) = (E - \hbar^2k^2/2m + i\eta)^{-1}$ corresponds to the free propagator and $\Sigma_{\ell j}$ is the reducible self-energy. 
The latter can be obtained by iterating the irreducible self-energy to all orders
\begin{eqnarray}\label{eq:redSigma1}
\Sigma_{\ell j}(k,k^\prime ;E)& = &\Sigma^*_{\ell j}(k,k^\prime ;E)\\ \nonumber
  &+&  \!\!       \int \!\! dq q^2 \Sigma^*_{\ell j}(k,q;E)G^{(0)}(q;E )\Sigma_{\ell j}(q,k^\prime ;E) .
\end{eqnarray}
We first concentrate on energies below the Fermi energy that involve quantities that elucidate properties of the ground state.
Below the Fermi energy the hole spectral function is determined by the imaginary part of the propagator
\begin{equation}
S_{\ell j}(k;E) = \frac{1}{\pi}\text{Im }G_{\ell j}(k,k; E) .
\end{equation}
For negative energies, the free propagator has no imaginary part and so according to Eq.~(\ref{eq:gdys1}) the spectral function reads
\begin{equation}
S_{\ell j}(k;E) = \frac{1}{\pi}G^{(0)}(k;E) \text{Im }\Sigma_{\ell j}(k,k;E)G^{(0)}(k;E) ,
\end{equation}
for energies where the imaginary part of the self-energy does not vanish.
The total spectral strength at $E$ for a given $\ell j$ combination, 
\begin{equation}
S_{\ell j}(E) = \int_{0}^\infty \!\!\! dk\ k^2\ S_{\ell j}(k;E) ,
\label{eq:specs}
\end{equation}
yields the spectroscopic strength per unit of energy.

The imaginary part of the CDBonn self-energy vanishes between the maximum energy of 1p2h and the minimum energy of 2p1h states.
Inside this domain, discrete solutions to the Dyson equation are obtained from the standard version of the Dyson equation 
\begin{eqnarray}
\lefteqn{G_{\ell j}(k, k^{\prime}; E) = \frac{\delta( k - k^{\prime})}{k^2}G^{(0)}(k; E) } 
\hspace{.5in} \label{eq:gdys2} & & \\
		& &	+ G^{(0)}(k; E) \int_0^\infty \!\!\! dq\ q^2\  \Sigma^*_{\ell j}(k,q; E)G_{\ell j}(q,k'; E) .
	\nonumber	
\end{eqnarray}
The solution for discrete poles utilizes the spectral representation of the propagator
\begin{eqnarray}
G _{\ell j}(k ,k' ; E) 
&=  \sum_m \frac{\bra{\Psi^A_0} a_{k\ell j}
\ket{\Psi^{A+1}_m} \bra{\Psi^{A+1}_m} a^\dagger_{k' \ell j} \ket{\Psi^A_0}
}{ E - (E^{A+1}_m - E^A_0 ) +i\eta }  & \nonumber \\
& +  \sum_n \frac{\bra{\Psi^A_0} a^\dagger_{k' \ell j} \ket{\Psi^{A-1}_n}
\bra{\Psi^{A-1}_n} a_{k \ell j} \ket{\Psi^A_0} }{
E - (E^A_0 - E^{A-1}_n) -i\eta} , &
\label{eq:prop}
\end{eqnarray}
where complete sets of states in the $A\pm1$ systems are inserted.
The continuum solutions in the $A\pm1$ systems are also implied in the completeness relations.
The numerators of the particle and hole components of the propagator represent the products of overlap functions associated with adding or removing a nucleon from the $A$-body ground state.
Following standard steps~\cite{Dickhoff08} the eigenvalue problem corresponding to Eq.~(\ref{eq:gdys2}) reads
\begin{eqnarray} 
\!\!\!\!\! \frac{ k^2}{2m} \phi^n_{\ell j}(k) +\!
  \int \!\! dq q^2
\Sigma^*_{\ell j}(k,q;\varepsilon^-_n)  &\phi^{n}_{\ell j}(q) & = 
\varepsilon^-_n \phi^{n}_{\ell j}(k) ,
\label{eq:DSeq}
\end{eqnarray}
where
\begin{equation}
\varepsilon^-_n=E^A_0 -E^{A-1}_n 
\label{eq:eig}
\end{equation}
for energies below the corresponding Fermi energy $\varepsilon^-_F=E^{A}_0-E^{A-1}_0$.
The notation
\begin{equation}
\sqrt{S_{\ell j}^n} \phi^n_{\ell j}(k) = \bra{\Psi^{A-1}_n}a_{k \ell j} \ket{\Psi^A_0} ,
\label{eq:overlap}
\end{equation}
for the overlap functions is introduced.
For an eigenstate of the Schr{\"o}dinger-like equation [Eq.~(\ref{eq:DSeq})], the so-called quasihole state labeled by $\alpha_{qh}$, the corresponding normalization or spectroscopic factor is given  by~\cite{Dickhoff08}
\begin{equation}
S^n_{\ell j} = \bigg( {1 - 
\frac{\partial \Sigma^*_{\ell j}(\alpha_{qh},
\alpha_{qh}; E)}{\partial E} \bigg|_{\varepsilon^-_n}} 
\bigg)^{-1} ,
\label{eq:sfac}
\end{equation}
which is the discrete equivalent of Eq.~(\ref{eq:specs}).
Discrete solutions in the domain where the self-energy has no imaginary part can therefore be obtained by expressing Eq.~(\ref{eq:DSeq}) on a grid in momentum space and performing the corresponding matrix diagonalization. 
This also applies to bound orbits above the corresponding Fermi energy given by $\varepsilon^+_F=E^{A+1}_0-E^A_0$.
We note that the DOM self-energy of Ref.~\cite{Charity07} contains an imaginary part below (hole domain) and above (particles) the average Fermi energy $\varepsilon_F =(\varepsilon^+_F + \varepsilon^-_F)/2$.

The momentum distribution for a given $\ell j$ is obtained from 
\begin{equation}
n_{\ell j}(k) = n^c_{\ell j}(k) + n^q_{\ell j}(k) ,
\label{eq:momdis}
\end{equation}
where the continuum contribution is obtained by integrating the spectral function up to corresponding threshold
\begin{equation}
n^c_{\ell j}(k) = \int_{-\infty}^{\varepsilon_T^-} dE\ S_{\ell j}(k;E) 
\label{eq:momd}
\end{equation}
and the contribution of the discrete quasihole states reads
\begin{equation}
n^q_{\ell j}(k) = \sum_n S^n_{\ell j} \left| \phi^n_{\ell j}(k)\right|^2 .
\label{eq:momdq}
\end{equation}
For protons the total momentum distribution (normalized by the number of protons $Z$) is obtained from
\begin{equation}
n(k) = \frac{1}{Z}\sum_{\ell j} (2j+1) n_{\ell j}(k) ,
\label{eq:tmomd}
\end{equation}
with a similar result for the neutron distribution.

Information about natural orbits can be generated by determining the one-body density matrix.
The  continuum contribution to the one-body density matrix reads 
\begin{equation}
n^{c}_{\ell j}(k^{\prime},k) = \frac{1}{\pi}\int_{-\infty}^{\varepsilon^-_T}dE\ S_{\ell j}(k, k^{\prime}; E)
\end{equation} 
where
\begin{equation}
S_{\ell j}(k,k^{\prime};E) = \frac{1}{\pi}G^{(0)}(k;E) \text{Im }\Sigma_{\ell j}(k,k^{\prime};E)G^{(0)}(k^{\prime};E)
\end{equation}
corresponds to the nondiagonal spectral density. The one-body density matrix also receives a contribution from the quasiholes according to
\begin{equation}
n_{\ell j}^{q}(k^{\prime},k) = \sum_n S^n_{\ell j}\phi^{n*}_{\ell j}(k)\phi^n_{\ell j}(k^{\prime}),
\end{equation}
where $S^n_{\ell j}$ is the spectroscopic factor and $\phi^n_{\ell j}(k)$ are the quasihole eigen functions in momentum space.  The total one-body density matrix is then given by
\begin{equation}
n_{\ell j}(k^{\prime},k) = n_{\ell j}^{q}(k^{\prime},k) +n_{\ell j}^{c}(k^{\prime},k) .
\label{eq:dmat}
\end{equation}
By diagonalizing the one-body density matrix given in Eq.~(\ref{eq:dmat}) one obtains the natural orbits for each $\ell j$ combination together with the corresponding occupation numbers.
It is therefore possible to write 
\begin{equation}
n_{\ell j}(k,k') = \sum_{i} n^{no}_{i \ell j} \phi^{no^*} _{i \ell j}(k) \phi^{no} _{i \ell j}(k')  ,
 \label{eq:norb}
\end{equation}
with $n^{no}_{i \ell j}, \phi^{no}_{i \ell j}(k)$ the corresponding occupation numbers and wave functions for natural orbit $i$.
We note that these wave functions are normalized to unity.
Results for quantities introduced in this section will be discussed in Sec.~\ref{sec:results} and wherever possible compared to the corresponding DOM quantities.

%
%
In the language of many-body theory, the elastic nucleon-nucleus scattering is determined by the on-shell matrix element of the reducible self-energy $\Sigma_{\ell j}(k,k^\prime ;E)$, since it is directly related to the $\mathcal{S}$-matrix by~\cite{Dickhoff08}
\begin{eqnarray}\label{Smatrix}
\bra{k_0}\mathcal{S}_{\ell j}(E)\ket{k_0}&\equiv&e^{2i\delta_{\ell j}} \\
&=&1-2\pi i\left(\frac{m k_0}{\hbar^2}\right)\bra{k_0}\Sigma_{\ell j}(E)\ket{k_0}, \nonumber
\end{eqnarray}
where $k_0=\sqrt{2 m E}/\hbar$, $m$ is the nucleon mass, and $E$ is 
the energy relative to the center-of-mass. The phase shift, $\delta_{\ell j}$, defined by Eq.~(\ref{Smatrix}) is in general a complex number.
Its real part yields the usual phase shift and its imaginary part is associated with the inelasticity of the scattering process and denoted by
\begin{equation}
\eta_{\ell j}=e^{-2\text{Im}(\delta_{\ell j})} .
\label{eq:inel}
\end{equation}
In general, the coupling to more complicated excitations in the self-energy implies a complex potential responsible for the loss of flux in the elastic channel, characterized by the inelasticities $\eta_{\ell j}$.

Because self-energy calculations at positive energy are rare, it is perhaps useful to include some relevant results in terms of the phase shifts $\delta_{\ell j}$ for the quantities that will be discussed later .
The scattering amplitude is given by
\begin{equation}
f_{m_s',m_s}(\theta , \phi ) = -\frac{4m\pi^2}{\hbar^2} \langle 
{\bm k}'m_s' | \Sigma(E) | {\bm k} m_s \rangle ,
\label{eq:23.16}
\end{equation} 
with wave vectors of magnitude $k_0$.
The matrix structure is usually represented by
\begin{equation}
[f(\theta,\phi)] = \mathcal{F}(\theta) I + \bm{\sigma} \cdot \hat{\bm{n}} 
\mathcal{G}(\theta) ,
\label{eq:23.17}
\end{equation}
based on rotational invariance and parity conservation.
The unit vector is given by $
\hat{\bm{n}}= \bm{k} \times \bm{k}'/|\bm{k} \times \bm{k}'|$,
and $\bm{\sigma}$ is formed by the Pauli spin matrices.
The relation between $\mathcal{F}$ and $\mathcal{G}$ and the phase shifts
determined by Eq.~(\ref{Smatrix}), can now be worked out yielding
\begin{eqnarray}
\mathcal{F}(\theta) &=&
\frac{1}{2ik} \sum_{\ell = 0}^{\infty}
\left[ (\ell +1)\left\{e^{2i\delta_{\ell +}}-1\right\} \right. \nonumber \\
&+& \left. \ell \left\{e^{2i\delta_{\ell -}}-1\right\} \right] P_\ell (\cos \theta )
\label{eq:23.17b}
\end{eqnarray}
and
\begin{equation}
\mathcal{G}(\theta)=
\frac{\sin \theta}{2k} \sum_{\ell = 1}^{\infty}
\left[ e^{2i\delta_{\ell +}}
-      e^{2i\delta_{\ell -}} \right] P_\ell' (\cos \theta ) .
\label{eq:23.17c}
\end{equation}
We employ the notation $\delta_{\ell \pm} \equiv \delta_{\ell j=\ell
\pm \frac{1}{2}}$ and $P_\ell'$ denotes the derivative of the Legendre 
polynomial with respect to $\cos \theta$.
The unpolarized cross section reads
\begin{equation}
\left(\frac{d\sigma}{d\Omega}\right)_{unpol}=|\mathcal{F}|^2 +
|\mathcal{G}|^2 .
\label{eq:23.17d}
\end{equation}
Employing the partial wave expansions~(\ref{eq:23.17b}) and (\ref{eq:23.17c})
and the orthogonality of the Legendre polynomials, we find
\begin{eqnarray}
\sigma^{el}_{tot} & = &
\frac{\pi}{k^2} \sum_{\ell = 0}^{\infty} \frac{
\left| (\ell +1)\left\{e^{2i\delta_{\ell +}}-1\right\}
+ \ell \left\{e^{2i\delta_{\ell -}}-1\right\} \right|^2}{2\ell +1} 
\nonumber \\
& + & \frac{\pi}{k^2} \sum_{\ell = 0}^{\infty} \frac{\ell (\ell +1)\left|
e^{2i\delta_{\ell +}}
-      e^{2i\delta_{\ell -}} \right|^2 }{2\ell +1} .
\label{eq:23.17h}
\end{eqnarray}
We can define partial elastic cross sections such that 
\begin{equation}
\sigma^{el}_{tot} = \sum_{\ell =0}^\infty \sigma^{el}_{\ell} ,
\label{eq:23.17z}
\end{equation}
which for a given $\ell$ read
\begin{equation}
\sigma^{el}_{\ell}= \frac{\pi}{k^2}\left[ (\ell +1)
\left| e^{2i\delta_{\ell +}}-1 \right|^2 + \ell \left|
e^{2i\delta_{\ell -}}-1 \right|^2 \right] .
\label{eq:23.17i}
\end{equation}
With complex potentials, and therefore complex phase shifts, it is possible
to calculate the total reaction cross 
section
\begin{equation}
\sigma^r_{tot} = \sum_{\ell =0}^\infty \sigma^{r}_{\ell} ,
\label{eq:23.17y}
\end{equation}
with
\begin{equation}
\sigma^{r}_{\ell}= \frac{\pi}{k^2}\left[ (2\ell +1) - (\ell +1)
\left| e^{2i\delta_{\ell +}}\right|^2 - \ell \left| 
e^{2i\delta_{\ell -}} \right|^2 \right] .
\label{eq:23.17k}
\end{equation}
These results are derived by using the optical theorem that yields the total 
cross section from the imaginary part
of the forward scattering amplitude~\cite{gott94}
\begin{equation}
\sigma_{T} = \sigma^{el}_{tot} + \sigma^{r}_{tot} .
\label{eq:23.17j}
\end{equation}
The results presented here refer to neutron scattering when the self-energy has a finite range.

Therefore, it is clear that at positive energies the problem is completely reduced to solving the integral equation for the reducible self-energy given in Eq.~(\ref{eq:redSigma1}).
It should be noted that the solution in momentum space automatically treats the non-locality of the reducible self-energy in coordinate space.
In practice, the integral equation is solved in two steps. 
First the integral equation is solved by only including the principal value part of the noninteracting propagator.
 Subsequently, it is straightforward to employ the resulting reaction matrix to take into account
the contribution of the $\delta$-function associated with the imaginary part of the noninteracting propagator.

Previous calculations of the nucleon self-energy for ${}^{16}$O~\cite{Muther94,Muther95,Polls95} only included a limited number of sp partial waves.
While this is sufficient for energies below the Fermi energy, results for differential cross sections at positive energy quickly require many more partial waves with increasing energy.
It is nevertheless possible to draw sensible conclusions by comparison with the corresponding partial wave contributions from DOM results that provide an accurate fit to the experimental data.

\section{Results}
\label{sec:results}
\subsection{Spectral Functions}
\label{sec:sfunc}

\begin{figure}[b]
\includegraphics[width=3.2in]{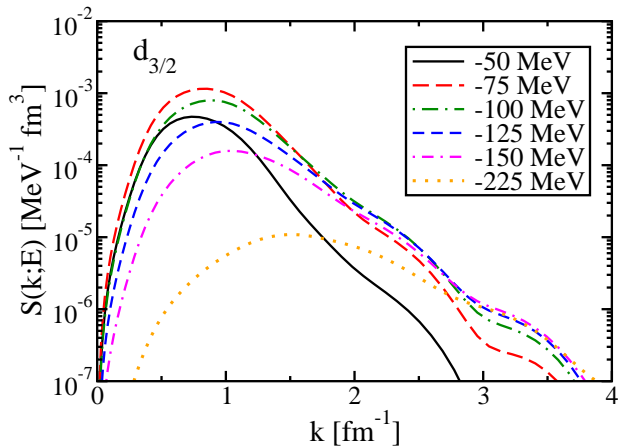}
\caption{Momentum-space spectral function for neutron $d_{3/2}$ quantum numbers at different energies in ${}^{40}$Ca.
\label{fig:d32sf} }
\end{figure}
In Fig.~\ref{fig:d32sf} the $d_{3/2}$ spectral function as a function of momentum is shown for different negative energies.
The $d_{3/2}$ orbit is the last one that is mostly occupied in ${}^{40}$Ca and the momentum  content of the state near the Fermi energy exhibits no substantial strength beyond the naive expectation, similar to what was found for the $p_{1/2}$ orbit in ${}^{16}$O before~\cite{Muther94}.
As illustrated in Fig.~\ref{fig:d32sf}, the strength at higher momenta increases with decreasing energy, as expected and consistent with earlier calculations~\cite{Muther95}.  
We note that the presence of the high-momenta is not as pronounced as observed previously in $^{16}$O in Refs.~\cite{Muther94,Muther95}. 
In this earlier work the Bonn-B potential~\cite{Machleidt89} was employed, whereas the present work employs the CDBonn interaction~\cite{cdbonn,cdbonna} which appears to be softer.
We note that the absence (presence) of high momenta near (far below) the Fermi energy is a simple consequence of simultaneous energy and momentum conservation and is well-documented for nuclear matter~\cite{ramos89}. 
The behavior of the spectral functions in Fig.~\ref{fig:d32sf} is somewhat different from the DOM results discussed in Ref.~\cite{Dickhoff10a}.
Whereas the microscopic strength exhibits a decrease of low momenta starting at -75 MeV, while the higher momenta exhibit similar strength, the corresponding DOM results show a more uniform behavior with little change in momentum profile as a function of energy.

\begin{figure}
\includegraphics[width=3.2in]{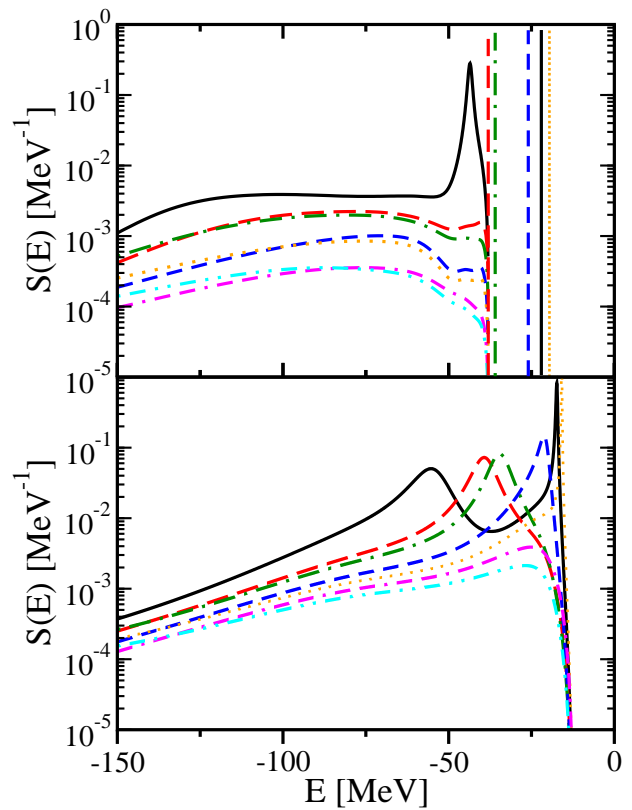}
\caption{ Comparison of spectral functions from present work (upper panel) with those from a DOM analysis using a non-local potential (lower panel). The results are for neutrons in $^{40}$Ca. The curves correspond to different $\ell j$-combinations that are identified in the text.
\label{fig:cdbdom}}
\end{figure}
In the top panel of Fig.~\ref{fig:cdbdom} the discrete and continuum contributions to the spectral strength [see Eq.~(\ref{eq:specs})] are shown as a function of the energy for various $\ell j$ channels for neutrons in $^{40}$Ca.
Results are shown for the following $\ell j$ combinations, $s_{1/2}$ (solid), $p_{3/2}$ (long-dash), $p_{1/2}$ (long-dash-dot), $d_{5/2}$ (dash), $d_{3/2}$ (dotted), $f_{7/2}$ (dash-dash-dot), and $f_{5/2}$ (dash-dot-dot).
In the bottom panel the corresponding DOM strength functions are shown for comparison.
The DOM self-energy from Ref.~\cite{Charity07} has an imaginary part that ends at $\varepsilon_{T}^- = \varepsilon_F$ and includes a surface term to account for long-range correlations (LRC).
The resulting strength distribution is therefore continuous with sharp peaks near the Fermi energy. These peaks are represented by discrete ones (normalized by spectroscopic factors in the figure) for the CDBonn self-energy.
Because the coupling to LRC is deemphasized, the imaginary part ends at a much lower energy $\varepsilon_{T}^- = -38$ MeV.
The DOM peaks closely correspond to the location of corresponding experimental sp states.  
The CDBonn self-energy generates energies for the peaks near the Fermi energy that are not too different from the DOM result.
For the deeply bound $s_{1/2}$ peak, the CDBonn result underestimates the binding by a substantial amount as compared to the DOM result which is consistent with a corresponding proton peak observed in the $(e,e'p)$ reaction~\cite{FM84}.

From Fig.~\ref{fig:cdbdom}, it is evident that the microscopic self-energy generates more sp strength at energies below the deeply-bound $s_{1/2}$ peak than the DOM self-energy used in Ref.~\cite{Dickhoff10a}.  
Another important difference is that the sp strength decreases more quickly with increasing $\ell$ in the microscopic approach than in the DOM.
We attribute this behavior to the non-locality of the imaginary part of the microscopic self-energy, whereas the imaginary part of the DOM self-energy is purely local. 
This feature has important consequences for the number of particles calculated from the corresponding spectral functions according to
\begin{equation}
N_{calc}=\sum_{\ell j}(2j+1)\int_{-\infty}^{\varepsilon^-_F} dE\ S_{\ell j}(E) .
\label{eq:ncalc}
\end{equation}
Since the microscopic self-energy shows a substantial reduction of sp strength with increasing $\ell$, the total particle number exhibits a good convergence with $N_{calc} = 19.3$ for neutrons when also the two partial waves with $\ell = 4$ are included.
The corresponding DOM result exhibits a much slower convergence as illustrated in Fig.~\ref{fig:cdbdom}, and crosses 20 already at the $f_{5/2}$ orbit~\cite{Dickhoff10a}.
We therefore conclude that the non-locality of the imaginary part of the self-energy plays an important role in generating the correct number of particles from a DOM self-energy.

\subsection{Quasiholes and quasiparticles}
The quasihole energies are shown in Table~\ref{tbl:qhE}. 
\begin{table}
\caption{Quasihole energies for neutron orbits in $^{40}$Ca. The second column shows the results from using the Hartree-Fock part only and third column the results from including the 2p1h and 2h1p terms in the self-energy. We also include DOM results and the position of the experimental sp levels near the Fermi energy.}
\label{tbl:qhE}
\begin{ruledtabular}
\begin{tabular}{ccccc}
orbit & BHF [MeV] & Full [MeV] & DOM [MeV] & Exp. [MeV] \\ \hline
$0s_{1/2}$ & -56.1 & -43.6 & -56.1 & $ $ \\
$0p_{3/2}$ & -37.4 & -33.9 & -39.6 & $ $ \\
$0p_{1/2}$ & -34.7 & -31.7 & -34.9 & $ $ \\
$0d_{5/2}$ & -20.4 & -21.8 & -21.6 & -22.3 \\
$1s_{1/2}$ & -18.1 & -19.6 & -17.4 & -18.3 \\
$0d_{3/2}$ & -16.0 & -17.8 & -15.9 & -15.6 \\
$0f_{7/2}$ &   -4.3        & -7.1 &      -9.8      &  -8.4 \\
$1p_{3/2}$ &     -2.6       &  -5.1    &      -7.0            &   -5.9      \\
$1p_{1/2}$ &        -1.2 &     -3.5         &      -5.4          &   -4.2  \\
\end{tabular}
\end{ruledtabular}
\end{table}
They were obtained by diagonalizing the self-energy in momentum space while taking the energy dependence of the real part of the self-energy properly into account but disregarding the imaginary part. 
We also include the result for the static contribution of the self-energy labeled by BHF in the table.
We note that the inclusion of the dispersive contribution moves the $0s_{1/2}$ state up by almost 13 MeV. 
The cause of this huge shift must be attributed to the strong energy dependence of the diagram (c) in Fig.~\ref{fig:diag} which is very repulsive at the solution of the $0s_{1/2}$ eigenvalue.
For other quasihole energies substantially smaller corrections of both signs are obtained. 
While not including the imaginary part of the self-energy does not yield the correct normalization for the $0s_{1/2}$ state, we do find that the energy in the table is consistent with the location of the corresponding peak in Fig.~\ref{fig:cdbdom}. 
Results for the DOM self-energy employed in Ref.~\cite{Dickhoff10a} are also included in the table as well as the experimental location of the sp orbits near the Fermi energy.
The particle-hole gap of the CDBonn self-energy is more than 10 MeV, substantially larger than for the DOM at 6.1 MeV which is a little below the experimental result of 6.8 MeV.
A common issue with microscopic self-energies is the underestimate of the spin-orbit splitting near the Fermi energy.
For the splitting of the $d$-states below the Fermi energy only 4 MeV is obtained compared to 6.7 MeV found experimentally.
The DOM generates 5.7 MeV for this quantity.
Relativistic effects and core polarization~\cite{zamick} or the importance of three-body forces~\cite{pieper} are usually invoked to repair this discrepancy but we note that only part of the sp strength resides in these levels, making the determination of the spin-orbit splitting more ambiguous.

\begin{table}[b]
\caption{Quasihole (quasiparticle) spectroscopic factors and occupation numbers for the CDBonn (CDB) self-energy compared to the corresponding DOM results.}
\label{tbl:sfocc}
\begin{ruledtabular}
\begin{tabular}{ccccc}
orbit & $S_{CDB}$ & $S_{DOM}$ & $n_{CDB}$ & $n_{DOM}$ \\ \hline
$1s_{1/2}$ & 0.85 & 0.66 & 0.91 & 0.88\\
$0d_{3/2}$ & 0.87 & 0.64 & 0.92 & 0.86\\
$0f_{7/2}$ & 0.92 & 0.67 & 0.02 & 0.11\\
$1p_{3/2}$ & 0.93 & 0.69 & 0.02 & 0.07\\
$1p_{1/2}$& 0.93 & 0.73 & 0.02 &0.06
\end{tabular}
\end{ruledtabular}
\end{table}
The spectroscopic factors identify the amount of sp strength residing near the Fermi energy and together with occupation numbers are shown in Table~\ref{tbl:sfocc} for the quasiparticle and quasihole states. 
The spectroscopic factors are calculated according to Eq.~(\ref{eq:sfac}) after solving the Dyson equation~(\ref{eq:DSeq}).
Occupation numbers for the quasiparticle or quasihole orbits are obtained by double folding the momentum space wave functions with the one-body density matrix in momentum space given in Eq.~(\ref{eq:dmat}).
For the hole states near $\varepsilon_F$ there is a reduction of sp strength of a little more than 10\% for the CDBonn calculation associated mostly with the effect of SRC. 
For the particle states, the reduction of the sp strength corresponds to about 10\%. 
It was observed in Ref.~\cite{Dickhoff10a} that the spectroscopic factors calculated for orbits in the continuum are not reliable (and can be larger than 1) so these are not included in the table.  
All CDBonn spectroscopic factors are about 20\% larger than in the DOM calculation. 
Since the DOM includes the coupling to low-lying excitations associated with collective effects of the nuclear surface, this simply reflects the important role of LRC that are responsible for this difference, and the resulting spectroscopic factors are in good agreement with the analysis of the $(e,e'p)$ reaction on ${}^{40}$Ca~\cite{Kramer89}.
Differences in integrated strength as displayed by occupation numbers are less dramatic partially because the strength removed to lower energy from the quasihole (particle) peaks due to LRC in the DOM is recovered in the occupation numbers~\cite{Charity11}.

\subsection{Momentum distribution}
\label{sec:momdis}
\begin{figure}[b]
\includegraphics[width=3.2in]{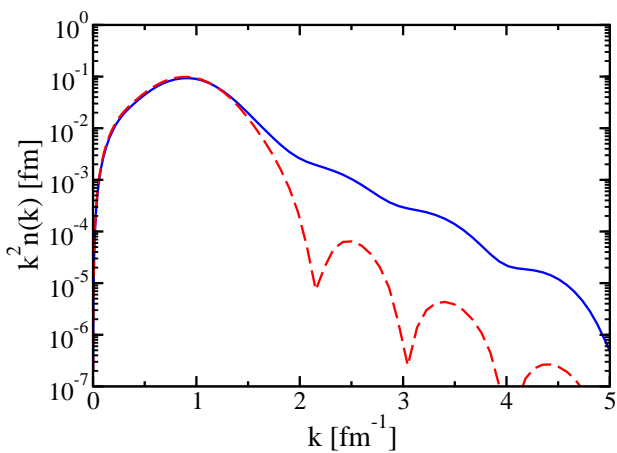}
\caption{(Color online) Momentum distribution for neutrons in $^{40}$Ca weighted by $k^2$. 
Solid line represents the total momentum distribution including quasihole and continuum terms.
The dashed line represents the quasihole result without reductions from spectroscopic factors.\label{fig:xmomdis}}
\end{figure}

\begin{figure}[b]
\includegraphics[width=3.2in]{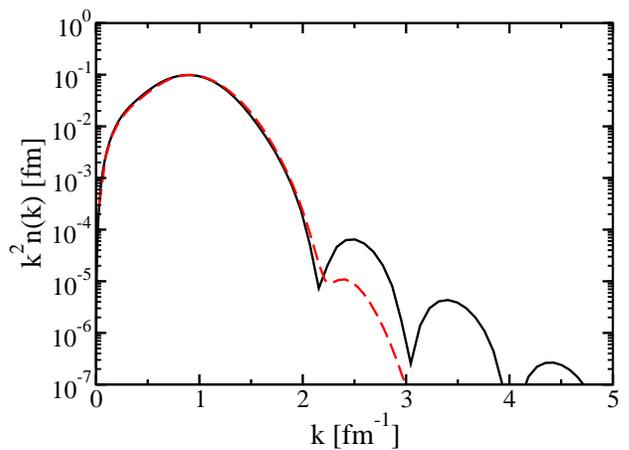}
\caption{(Color online) Quasihole contribution to the momentum distribution for the CDBonn (solid) compared to the DOM result (dashed). Both distributions are normalized according to $4\pi \int dk\ k^2 n(k) =1$.\label{fig:qhmd}}
\end{figure}

The total momentum distribution for neutrons in $^{40}$Ca resulting from the CDBonn interaction and calculated according to the equivalent of Eq.~(\ref{eq:tmomd}) is shown in Fig.~\ref{fig:xmomdis} by the solid line.
All distributions are multiplied by $k^2$ and normalized such that $4\pi \int dk\ k^2 n(k) =1$.
The quasihole contribution without the reduction due to spectroscopic factors is shown by the dashed line.
Comparison of the correlated with this mean-field-like contribution (dashed) shows an appreciable presence of high-momentum components. 
\begin{figure}[t]
\includegraphics[width=3.2in]{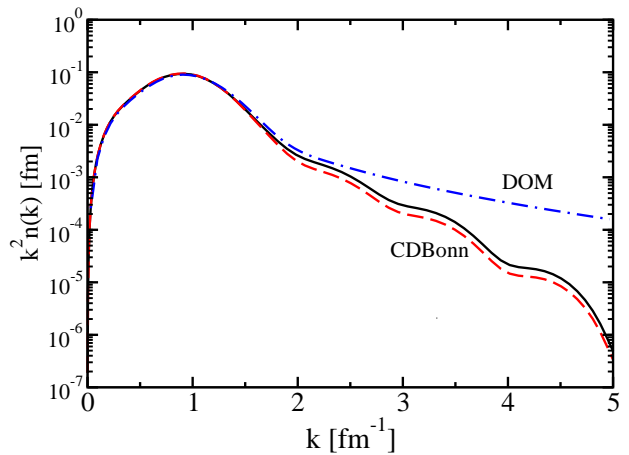}
\caption{(Color online) CDBonn momentum distribution with $\ell_{max}=3$ (dashed) and $\ell_{max}=4$ (solid) compared with the DOM result including all partial waves including $f_{7/2}$ (dash-dot) as obtained in Ref.~\protect{\cite{Dickhoff10a}}. 
The normalization of the curves is given by $4\pi \int dk\ k^2 n(k) =1$\label{fig:cdbDOM}}
\end{figure}
A comparison with the DOM result (dash-dot) in Fig.~\ref{fig:cdbDOM}, normalized as in Fig.~\ref{fig:xmomdis}, indicates that the DOM self-energy has slightly more high-momentum components than the microscopic one, although  the latter has a larger quasihole contribution at high-momentum, as shown Fig.~\ref{fig:qhmd}. 
The amount of strength above 1.4 fm$^{-1}$ corresponds to 8\% for the CDBonn calculation compared to 10\% for the DOM.
The convergence of the CDBonn result with orbital angular momentum is also illustrated in Fig.~\ref{fig:cdbDOM} exhibiting a satisfactory convergence when the $\ell_{max} =3$ result (dashed) is compared with the one for $\ell_{max}=4$ (solid).
This convergence is less satisfactory for the DOM result~\cite{Dickhoff10a} which contains contributions from partial waves only up to and including the $f_{7/2}$ one.
These results further confirm the importance of a non-local representation of the imaginary part of the self-energy which automatically leads to a more satisfactory convergence with orbital angular momentum.
As discussed in Sec.~\ref{sec:sfunc}, the energy dependence of the spectral function of the CDBonn potential in momentum space already suggests that it is a rather soft potential in comparison with the Bonn-B potential~\cite{Machleidt89} that was employed for ${}^{16}$O in Refs.~\cite{Muther94,Muther95,Polls95}.
While the spectroscopic factors for the aforementioned Bonn potentials in these nuclei are similar, the CDBonn potential contains about 4\% of strength in the quasihole orbits above 1.4 fm$^{-1}$, whereas for the Bonn-B potential this amount is much smaller.
It would be interesting in future to investigate the corresponding behavior of a harder and local interaction such as the Argonne $V18$~\cite{wiringa95}. 

\subsection{Natural Orbits}
\label{sec:natorb}

\begin{center}
\begin{table*}[bth] {
\caption{Occupation numbers of natural orbits}
\label{tbl:no}
\begin{ruledtabular}
\begin{tabular}{cccccccc}
n & $s_{1/2}$ & $p_{3/2}$ & $p_{1/2}$ & $d_{5/2}$ & $d_{3/2}$ & $f_{7/2}$ & $f_{5/2}$ \\ \hline
1 & 0.882 & 0.902 & 0.898 & 0.909 & 0.919 & 0.024 & 0.025\\
2 & 0.910 & 0.025 & 0.025 & 0.014 & 0.015 & 0.006 & 0.007 \\
3 & 0.016 & 0.007 & 0.008 & 0.003 & 0.004 & 0.001 & 0.001 \\
4 & 0.004 & 0.001 & 0.002 & 0.001 & 0.001 & 0.0003 & 0.0005 \\
5 & 0.001 & 0.0004 & 0.0004 & 0.0001 & 0.0002 & 0.0001 & 0.0001 \\
6 & 0.0002 & 0.0001 & 0.0001 & 0.0001 & 0.0001 & $<$ 1e-4  & $<$ 1 e-4 \\
$\Sigma_n$ & 1.82 & 0.94 & 0.93 & 0.93 & 0.94 & 0.03 & 0.03 \\
\end{tabular} 
\end{ruledtabular} }
\end{table*}
\end{center}

\begin{figure}[b]
\includegraphics[width=3.2in]{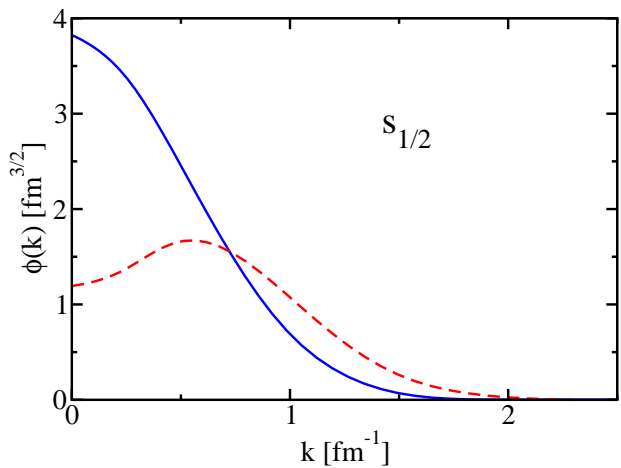}
\caption{(Color online) Comparison of wave functions for the $s_{1/2}$ quasihole result (solid) and the corresponding natural orbit (dashed) without a node.   \label{fig:0s} }
\end{figure}
\begin{figure}
\includegraphics[width=3.2in]{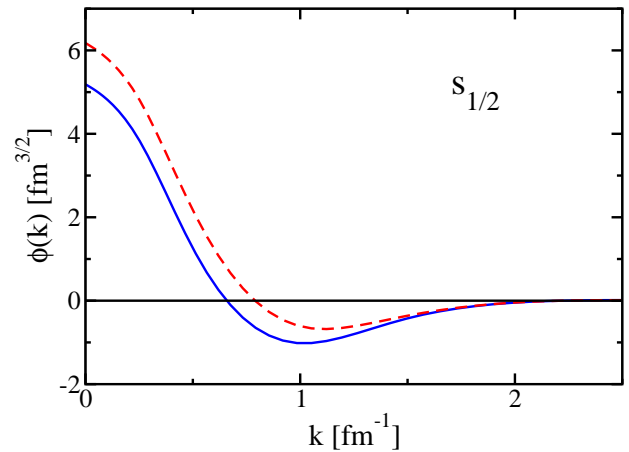}
\caption{ As in Fig.~\ref{fig:0s} but showing $s_{1/2}$ wave functions with one node. \label{fig:1s}  }
\end{figure}
\begin{figure}[b]
\includegraphics[width=3.2in]{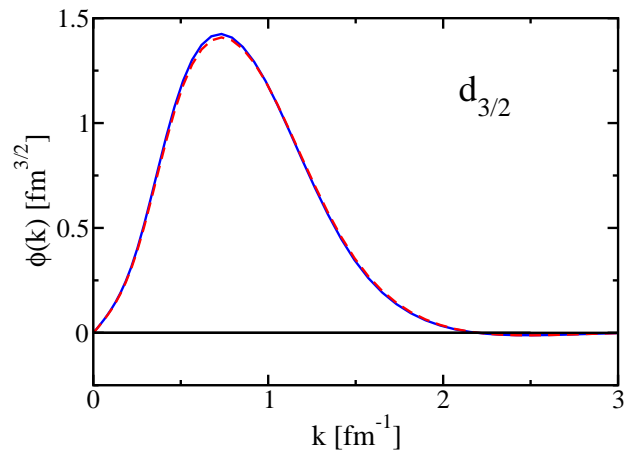}
\caption{ As in Fig.~\ref{fig:0s}, but with the corresponding $d_{3/2}$ wave functions. \label{fig:0d}}
\end{figure}

Results for natural orbits are obtained by diagonalizing the one-body density matrix which leads to Eq.~(\ref{eq:norb}) containing the natural orbit functions in momentum space (shown in Figs.~\ref{fig:0s}-\ref{fig:0d}) accompanied by the corresponding occupation numbers.
In Table~\ref{tbl:no} results for the occupation numbers of natural orbits are collected for the relevant partial waves. 
The results are quantitatively similar to previous results for $^{16}$O~\cite{Polls95}.
More surprisingly, the results for the occupation numbers of the natural orbits that are predominantly occupied are very close to the DOM result reported in Ref.~\cite{Dickhoff10a}.
Only the orbits immediately adjacent to the Fermi energy exhibit more than 1 or 2\% deviation but never more than 8\%.
This is perhaps surprising in light of the rather large difference between the spectroscopic factors obtained with these different approaches.
Since both approaches are constrained by very different experimental data, it is therefore gratifying that the dominant occupation numbers for the natural orbits are so close.
Differences of a few percent are observed for the smaller occupation numbers leading to a total occupation of these DOM orbits that is a few percent higher than in the case of the CDBonn.
We ascribe this to the inclusion of LRC in the DOM.

A comparison with natural orbits obtained for finite drops of ${}^3$He atoms~\cite{Lewart88} illustrates the substantial difference between the underlying fermion-fermion interactions.
The atom-atom interaction being much more repulsive and in a larger domain, relative to the volume per particle, leads to occupation numbers as small as 0.54 for the $1s$ state in a drop of 70 ${}^3$He atoms, whereas the nuclear interaction generates values close to 0.9 either in the DOM or from the CDBonn calculation.
The difference is therefore mostly related to the much stronger repulsion between ${}^{3}$He atoms which \textit{e.g.} in the liquid at saturation leads to a depletion of the Fermi sea of more than 50\%~\cite{Mazz04}. 
Nucleon-nucleon interactions typically generate 10-15\% depletion due to SRC~\cite{Rios09}.

Wave functions of natural orbits and quasihole states are compared in Figs.~\ref{fig:0s}-\ref{fig:0d}.
In Fig.~\ref{fig:0s} the momentum-space wave functions without a node for the $s_{1/2}$ partial wave are shown for the deeply bound quasihole (solid) and the corresponding natural orbit (corresponding to an occupation number of 0.882).
The natural orbit wave function extends farther out in momentum space reflecting the contribution from lower energies that contain higher momenta.
The quasihole wave function for the $s_{1/2}$ state near the Fermi energy (solid) is compared to the natural orbit result (also with one node) in Fig.~\ref{fig:1s}.
Again there is a substantial difference between the two wave functions although they both have the same occupation number.
Contrary to the two $s_{1/2}$ results, the $d_{3/2}$ quasihole and natural orbit wave functions are essentially indistinguishable, as shown in Fig.~\ref{fig:0d}.
A similar result was obtained (in coordinate space) for the DOM calculation of Ref.~\cite{Dickhoff10a} and appears to be due to the presence of only one natural orbit with a large occupation number unlike the $s_{1/2}$ case.
We note that in the DOM calculation of Ref.~\cite{Dickhoff10a} both quasihole $s_{1/2}$ wave functions were essentially identical to the natural orbit results.

\subsection{Charge Distribution}
\label{sec:charged}

\begin{figure}[b]
\includegraphics[width=3.2in]{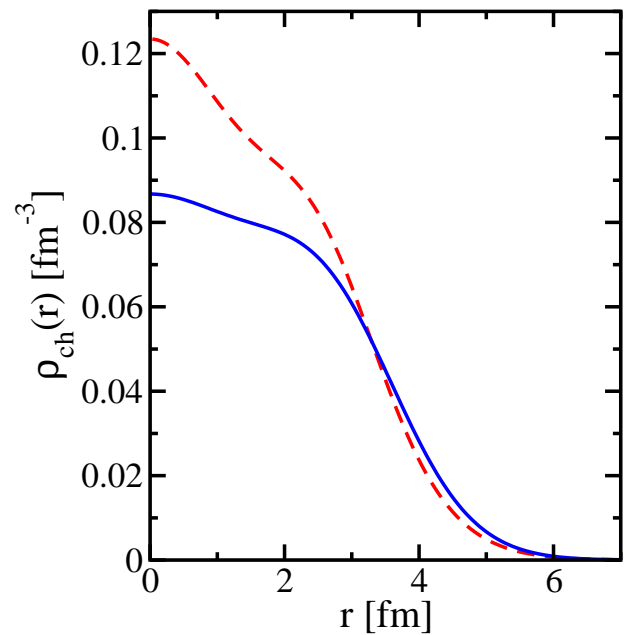}
\caption{ Charge density distribution for ${}^{40}$Ca from the CDBonn self-energy (dashed) compared to experiment (solid). \label{fig:cd}}
\end{figure}

Although we focus mainly on neutron results for ${}^{40}$Ca, it is useful to study the charge density distribution obtained for the CDBonn potential.
For this purpose, the Coulomb potential was incorporated into the calculations by first transforming the irreducible self-energy to coordinate space. 
The Coulomb potential from a uniformly charged sphere was then added, and  a matrix inversion was performed to get the propagator in a similar way as in Ref.~\cite{Dickhoff10a}. 
The radius of the sphere was taken to be $R_C=1.31 A^{1/3}$, which is what was used in the DOM analysis. 
The point charge distribution was then obtained from
\begin{equation}
\rho_{ch}(r)=\frac{e}{4\pi}\sum_{\ell j}(2j+1)n_{\ell j}(r,r) ,
\end{equation}
where $n_{\ell j}$ refers to the one-body density matrix in coordinate space. 
The resulting distribution
was then corrected for the experimental charge distribution of a single proton and a single neutron as in Ref.~\cite{Brown79}. 
The final charge distribution is shown by the dashed line in Fig.~\ref{fig:cd} and compared to the experimental one obtained from the Fourier-Bessel analysis of Ref.~\cite{deVries1987}.

The mean square radius of the CDBonn self-energy is obtained from
\begin{equation}
\label{eq:msr}
\langle r^2 \rangle = \frac{1}{Ze} \int_0^\infty \!\!\!\! dr\ r^2 \rho_{ch}(r) ,
\end{equation}
yielding a value of 3.29 fm compared to the experimental result of 3.45 fm taken from Ref.~\cite{deVries1987}.
We note that microscopic calculations usually underestimate the experimental results (see \textit{e.g.} Ref.~\cite{Muther95} for ${}^{16}$O).
Since LRC correlations are not adequately incorporated, it is possible that an improved charge density is obtained when they are included.
We note that a recent microscopic calculation of the matter density in ${}^{40}$Ca also concentrates too much matter near the origin~\cite{Hagen08} suggesting that it is useful to consider comparison with as many experimental quantities as possible for a more detailed understanding of the quality of the many-body calculations.

We note that the DOM charge density also contains too much charge near the origin~\cite{Dickhoff10a} even though LRC are incorporated.
The DOM self-energy was however constrained to reproduce the mean-square-radius of the charge distribution.
It appears therefore essential to include apart from SRC, as shown here for the CDBonn, both non-locality and LRC in the DOM to obtain a charge density that is in better agreement with the data.
Indeed, future fits to experimental data in the DOM framework can utilize elastic electron scattering data directly to fit the parameters for nuclei where such data are available.

\subsection{Energy of the ground state}
\label{sec:energy}

From the momentum distribution and the spectral function in $k$-space, the neutron contribution to the ground state per neutron can be calculated by using: 
\begin{eqnarray}
\frac{E_n({}^{40}\textrm{Ca})}{N_{calc}}= \frac{1}{2N_{calc}}\sum_{\ell j}(2j+1)\int dk k^2 \frac{\hbar^2 k^2}{2m}
                                         n_{\ell j}(k) \nonumber  \\
                                       +\frac{1}{2N_{calc}}\sum_{\ell j}(2j+1)\int dk k^2 \int_{-\infty}^{\varepsilon_F} dE\ E\ S_{\ell j}(k;E) ,  \nonumber \\ \label{eq:epern}
\end{eqnarray}
where $E_n$ is the total energy from the neutrons, and $N_{calc}$ [see Eq.~(\ref{eq:ncalc})] is the calculated number of neutrons when all partial waves with $\ell \le 4$ are included. 
With this limit on the number of partial waves $N_{calc}=19.3$ and a corresponding energy per neutron of -8.25 MeV is obtained. 

As for the charge distribution calculation, the energy for the protons was generated in coordinate space. The density matrix and spectral function were then transformed back to the momentum space in order to use Eq. (\ref{eq:epern}) with $E_n$ replaced by $E_p$ and $N_{calc}$ by the calculated proton number $Z_{calc}$, which is found to be $Z_{calc}=19.5$. The resulting energy per proton is --4.91 MeV, and the resulting total energy per particle is
$E({}^{40}\textrm{Ca})/A$ = --6.56 MeV. This result is 1.85 MeV per particle more attractive than the DOM result in Ref.~\cite{Dickhoff10a}, but still more than 2 MeV/A higher than the experimental binding of --8.55 MeV/$A$. 
From the results of the spectral functions it is clear that more strength occurs in the continuum for the CDBonn at very negative energies than in the DOM calculation demonstrating the importance of these continuum contributions to the total energy.

This importance was also recognized in Ref.~\cite{Muther95}, where it was shown that about two-thirds of the binding is due to the continuum even though it represents only about 10\% of the particles.
Since a different interaction is used and a heavier nucleus is considered in the present work, it is instructive to quantify the role of the continuum also for the CDBonn calculation.
The quasihole contribution to the energy of the ground state are well separated from the continuum except for the peak corresponding to the lowest $s_{1/2}$ orbit as shown in Fig.~\ref{fig:cdbdom}.
We therefore assessed the $s_{1/2}$ continuum contribution by integrating the strength in Eq.~(\ref{eq:epern}) up to --50 MeV.
The total binding from the continuum contributions of all partial waves then amounts to --105.88 MeV compared to a total of --159.16, therefore representing 67\% of the total, a very similar result to the ${}^{16}$O calculations of Ref.~\cite{Muther95}.
We note further that even though the CDBonn interaction is relatively soft, the cancellation between kinetic energy (419.15 MeV) and potential energy (--578.32 MeV) is quite substantial.

Recent calculations employing the unitary-model-operator approach generate --8.51/$A$ close to experiment for ${}^{40}$Ca also using the CDBonn interaction~\cite{Fujii09}. 
The present calculation of --6.56 MeV/$A$ is almost 2 MeV per particle less, pointing to the need of additional correlations and an improved treatment of the propagators included in the present self-energy calculation.
We note that only noninteracting propagators have been used in the construction of the self-energy and that self-consistency was not attempted.
In addition, the proper treatment of LRC may also be relevant for certain observables like the charge density.
A microscopic treatment of LRC is available for the self-energy in which random-phase-approximation phonons are summed to all orders in a Faddeev technique and inserted into the self-energy~\cite{Barb:1,Barb:2,Barbieri07,Barb09,Barbieri09a,Degroote11}.
This FRPA method does not explicitly include high-momentum components so a combination of the current method and the FRPA needs to be developed.  
Self-energy calculations also have the additional advantage that other observables of interest can be evaluated.
Such observables are provided in particular by elastic-nucleon scattering data.
We turn to an analysis of the corresponding results from the CDBonn self-energy in the following section.

\subsection{Neutron-${}^{40}$Ca scattering}
\label{sec:scattering}
\begin{figure}
\includegraphics[width=3.4in,clip=true]{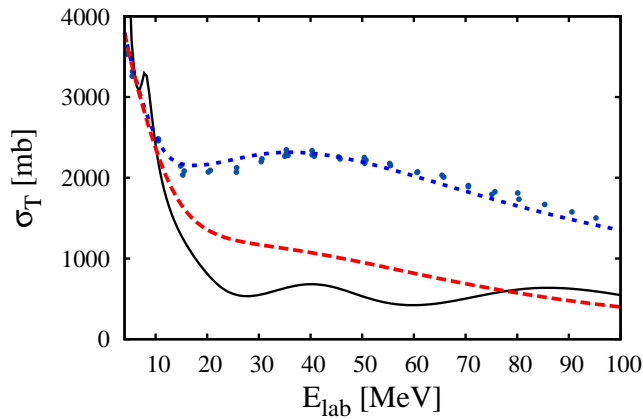}
\caption{ (Color online) Comparison of the experimental cross section (data points) with the CDBonn calculation (solid line) for n$+ ^{40}$Ca. The dotted line corresponds to the DOM fitted cross section, and the dashed line represents the DOM result including up to $\ell=4$ only.  The cross section obtained from the CDBonn self-energy has contributions only up to $\ell=4$.}
\label{sigmaTot1}
\end{figure}

In this section we make a direct comparison between CDBonn and DOM results, and experimental data for neutron-nucleus scattering, more specifically the $n+^{40}$Ca process.
References to data can be found in Ref.~\cite{Charity07,Charity11}.
The total cross section obtained from CDBonn self-energy is comparable to the experimental data
for energies between 5 to 10 MeV, while for larger energies it does not 
reproduce the observations as shown by the solid line in Fig.~\ref{sigmaTot1}. 
However, currently our CDBonn self-energy only includes contributions up to $\ell_{max}=4$ which is a reasonable limit for calculating properties below the Fermi energy. 
If we consider the DOM potential and only add the partial-wave contributions to the total cross section up to $\ell_{max}=4$, the corresponding result also fails to reproduce the experimental data, as shown by the dashed line in Fig. \ref{sigmaTot1}.  
The most useful procedure is therefore to compare the DOM and CDBonn calculation with the inclusion of the same number of partial waves.
It is then gratifying to observe that the microscopic calculation generates similar total cross sections as the DOM for energies where surface contributions are expected to be less relevant, \textit{i.e.} above 70 MeV.   
Although calculations for more partial waves are in principle possible, the amount of angular momentum recoupling corresponding to the appropriate basis transformations becomes increasingly cumbersome.
Nevertheless, it is clear that the current limit of $\ell_{max}=4$ is insufficient to describe the total cross section already at relatively low energy irrespective of whether surface effects (LRC) are properly included.

\begin{figure}[t]
\includegraphics[totalheight=4.3in,clip=true]{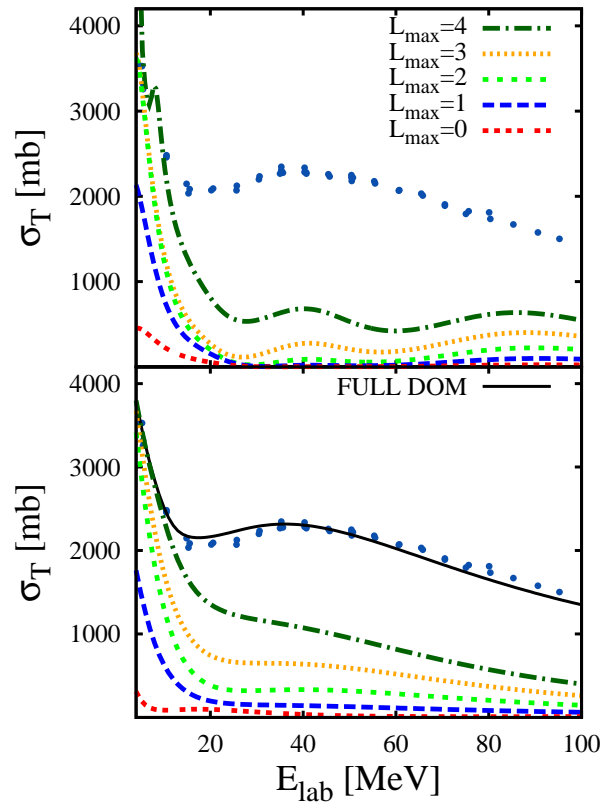}
\caption{(Color online) Effect of adding subsequent partial waves for the total cross section from the CDBonn self-energy (upper panel) and corresponding DOM result (lower panel). The total DOM cross section with a converged contribution of the partial-wave decomposition is also shown as the solid line in the lower panel.}
\label{sigmaTot2}
\end{figure}

To visualize in more detail the specific contribution from each partial wave, in Fig.~\ref{sigmaTot2}  
we display cross sections calculated up to a specific $\ell_{max}$ for CDBonn (upper panel) and DOM (lower panel). Despite the limitation of not having CDBonn values for $\ell>4$, both potentials provide a comparable cross sections up to this angular momentum cut-off in the explored energy range.
In the lower panel we also include the converged cross section of the DOM potential that provides a good description of the data.

\begin{figure}
\includegraphics[width=3.in]{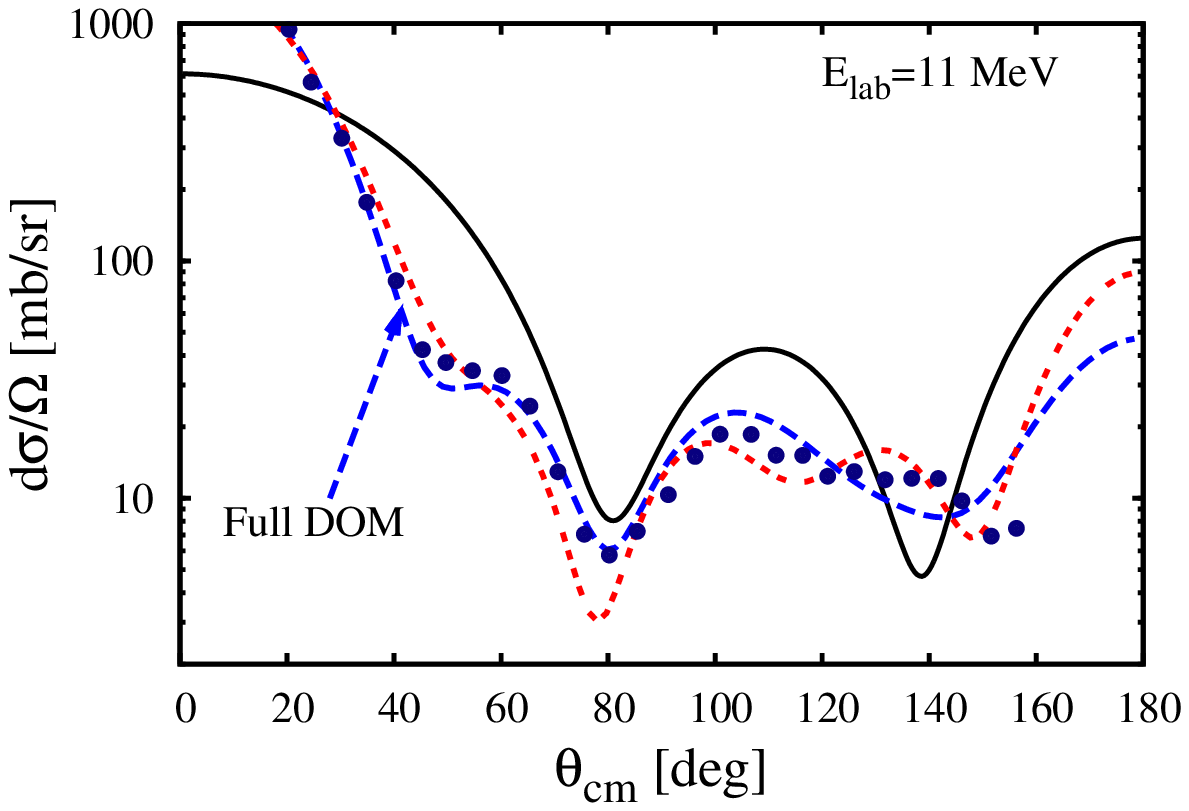}
\includegraphics[width=3.in]{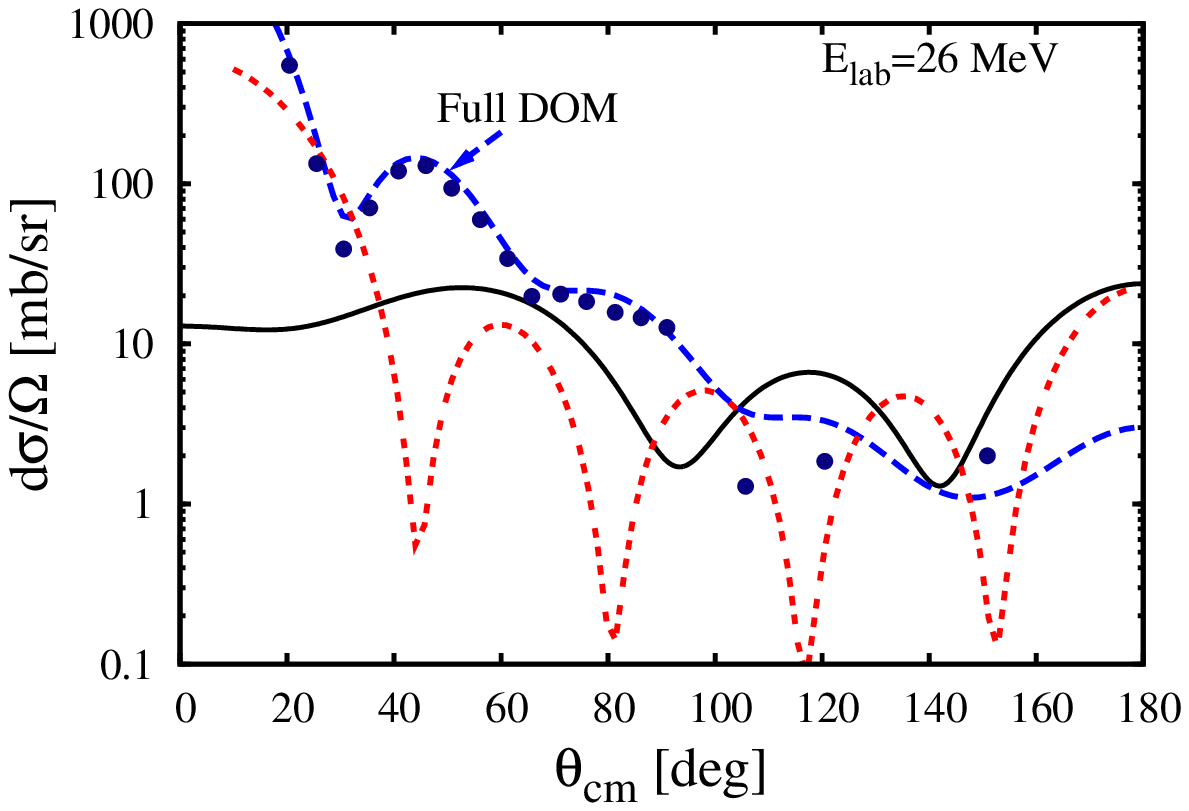}
\includegraphics[width=3.in]{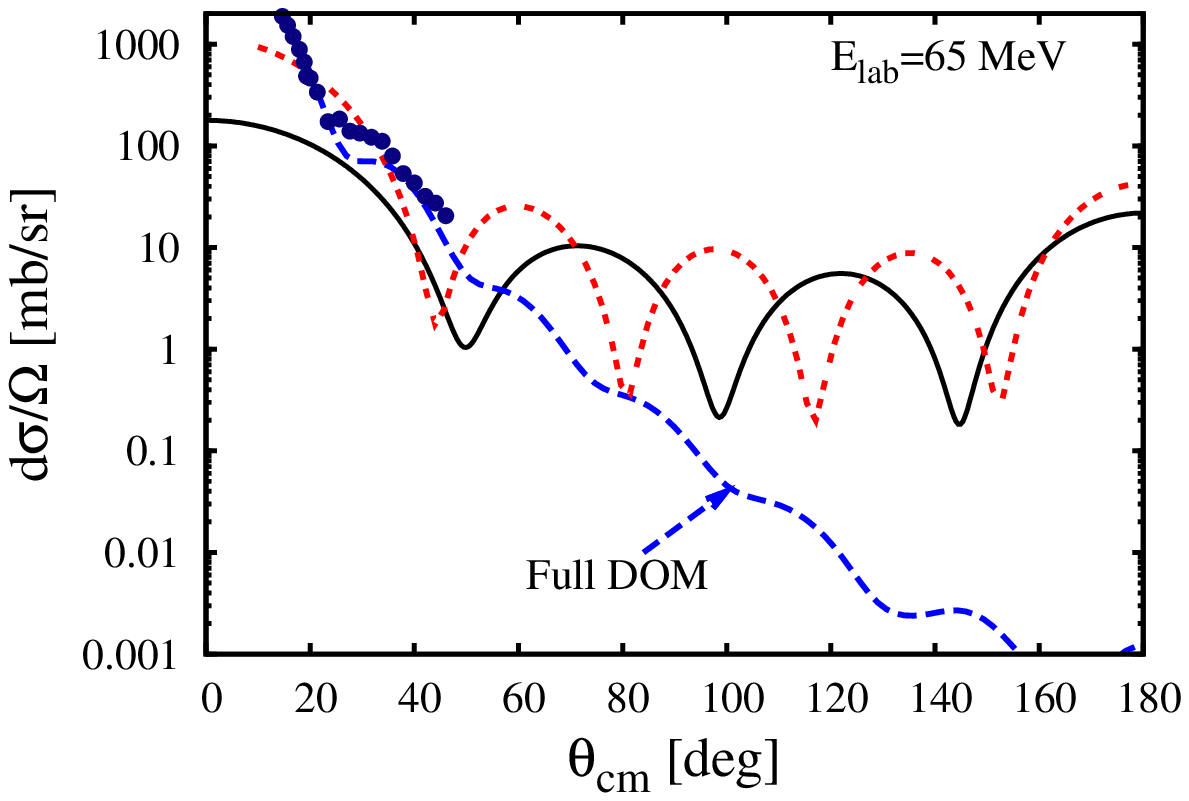}
\includegraphics[width=3.in]{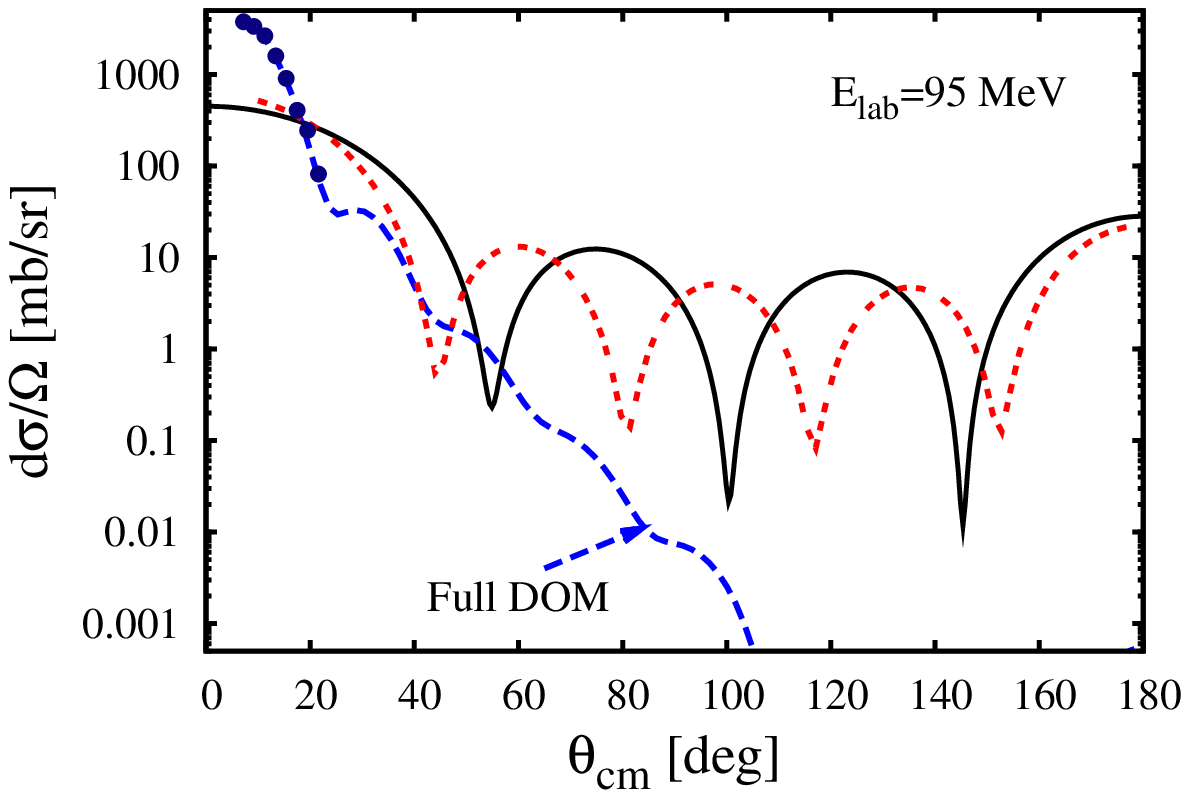}
\caption{(Color online) Differential cross section for indicated energies. Shown are the CDBonn result (solid), DOM for the same $\ell_{max}=4$ (short-dashed), and the full DOM cross section (dashed).
}
\label{ds_dw}
\end{figure}
In Fig.~\ref{ds_dw}, each panel displays the differential cross section as a function of the scattering angle for the indicated energies, covering 11, 26, 65, and 95 MeV for which data are available (shown by the dots). The solid line in every panel corresponds to the CDBonn result and the others to the DOM, with the dashed line representing the converged cross section and the short-dashed line limiting the sum up to $\ell_{max}=4$. 
It is quite sobering to realize how far from the experimental data the microscopic calculations are, especially at the lower energies.
This issue represents an important challenge to all many-body approaches that are of interest to the study of rare isotopes, since only hadronic reactions are available to study these and their analysis relies on presently unavailable optical potentials.
At the higher energies (65 and 95 MeV) the differential cross section, while somewhat out of phase with the DOM result (up to $\ell=4$), is of similar magnitude.
Clearly, the missing details of LRC at lower energy are needed to get better agreement with the data, while the missing $\ell$-contributions also become quickly noticeable.
Future work clearly requires the  calculation for higher $\ell$-values, but should also focus on an improved inclusion of surface dynamics in order to better reproduce the experimental nucleon-nucleus scattering data.

\begin{figure}
\includegraphics[width=3.in]{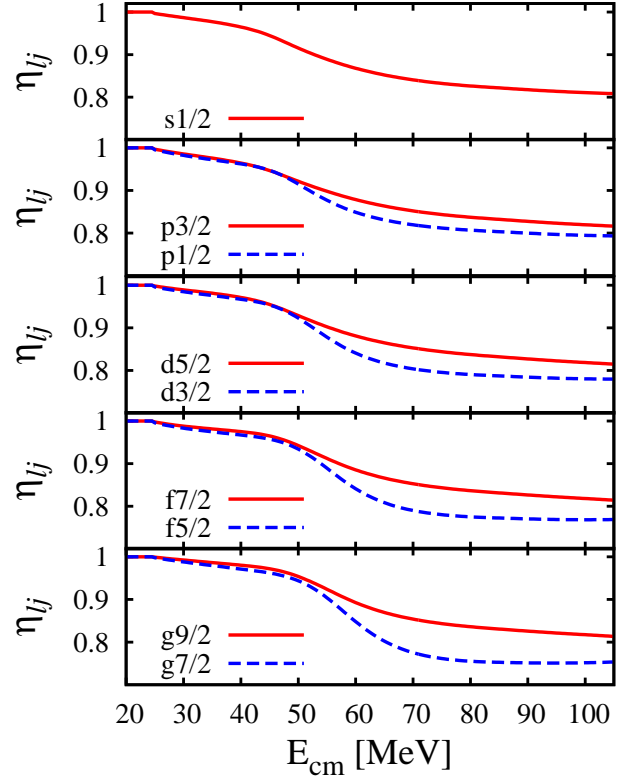}
\caption{ (Color online) In each panel, the inelasticity $\eta_{\ell j}$ for the indicated states as a function of the centre of mass energy. The solid (dashed) lines correspond to $j=| \ell+s |$ ($j=| \ell-s |$) states. Below 25 MeV, the CDBonn potential is real, therefore the $\eta_{\ell j}$ are equal to one. Above 25 MeV the decrease of the $\eta_{\ell j}$ indicates a  loss of flux as the energy increases.}
\label{inelasticities}
\end{figure}

In order to gain more insight into the similarities and differences between the CDBonn self-energy and DOM potentials, we also display the inelasticities [see Eq.~(\ref{eq:inel})].
The CDBonn's inelasticities $\eta_{\ell j}$ for each partial wave are presented 
in Fig.~\ref{inelasticities} as a function of energy. 
As expected $\eta_{\ell j}$ is equal to 
1 for energies below 25 MeV since the CDBonn potential is purely real from
$-38$ to 25MeV. 
We observe that the loss of flux above 25 MeV for each partial wave as 
a function of energy is larger for $\ell^+$ states than for $\ell^-$ ones
starting at energies $\simeq 50$ MeV, which reflects the presence of a non-negligible spin-orbit content in the imaginary part of the CDBonn self-energy. 
As expected, this difference is more pronounced for higher orbital angular momenta. In the range of energies considered all the inelasticities are always larger than 0.7.

\begin{figure}
\includegraphics[width=3.in]{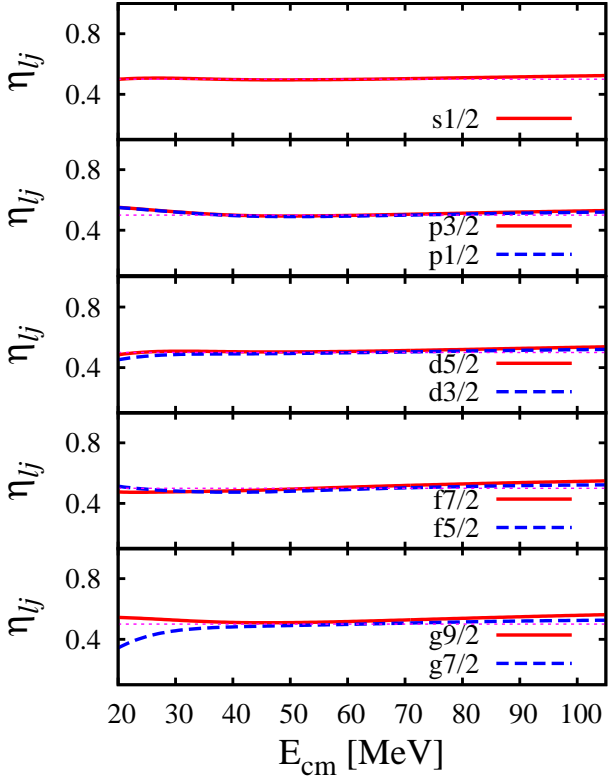}
\caption{ (Color online) In each panel, the inelasticity $\eta_{\ell j}$ for the indicated states as a function of the centre of mass energy obtained from the DOM potential. The solid (dashed) lines correspond to $j=| \ell+s |$ ($j=| \ell-s |$) states. The short dotted line indicates 0.5, note that the DOM potential predicts inelasticities about 0.5 without significant variations with energy or orbital angular momentum. }
\label{etasDOM_lj}
\end{figure}

We contrast these results with the inelasticities derived from the DOM potential for the same partial waves, as shown in Fig.~\ref{etasDOM_lj}.
The inelasticities derived from the DOM potential display little variation in the considered energy range, and saturate quickly and relatively close to 0.5.
As anticipated the DOM potential is therefore more absorptive than the CDBonn self-energy. This difference can be attributed to the lack of surface dynamics of the latter.
We also note the absence of an appreciable imaginary spin-orbit potential in the DOM potential, in contrast with the CDBonn result.
It is unclear whether the absence of an imaginary spin-orbit potential in the DOM is due to a compensating contribution that is not present in the current CDBonn self-energy.
It is also unclear without further analysis whether the apparent absence of a need for a non-local potential in the analysis of elastic scattering is due to the truly local character of the self-energy or due to a conspiracy of other contributing factors.
We perform some initial analysis into this issue in the next section.

\subsection{Analysis of the CDBonn self-energy\label{sec:analysis}}

Improving the analysis of elastic scattering data above the Fermi energy and observables related to quantities below the Fermi energy in a DOM framework appears to depend sensitively on the treatment of non-locality in the imaginary part of the self-energy.
It is therefore useful to gain some insight into the properties of microscopic self-energies which may offer guidance how to implement such features in the future.
We therefore performed a few simple fits to represent the central part of the imaginary part of the  CDBonn self-energy in coordinate space at a given energy assuming the following form of the potential
\begin{equation}
W_{NL}(\bm{r},\bm{r}') = W_0 \sqrt{f(r)}\sqrt{f(r^\prime)}H\left(\frac{\bm{r} - \bm{r}^\prime}{\beta}\right) .
\label{eq:nlocal}
\end{equation}
We deviate from the standard Perey prescription for non-locality by employing square-root factors of the function $f(r)$ which is still represented by the conventional Woods-Saxon form factor 
\begin{equation}
f(r)=\frac{1}{1+e^{\frac{r-R}{a_0}}} ,
\label{eq:WS}
\end{equation}
with $R=r_0 A^{1/3}$ .
The function $H$ determines the degree of non-locality and is assumed to be a gaussian following Ref.~\cite{Perey62}
\begin{equation}
H\left(\frac{\mathbf{r} - \mathbf{r}^\prime}{\beta}\right) = \frac{1}{\pi^{3/2}\beta^3}\text{exp}\left(\frac{|\mathbf{r} - \mathbf{r}^\prime|^2}{\beta^2}\right) .
\end{equation}
When the angular dependence in $H$ is projected out, an analytic solution is obtained for each orbital angular momentum $\ell$
\begin{eqnarray}
W^{\ell}_{NL}(r,r') &=& W_0 \sqrt{f(r)}\sqrt{f(r^\prime)}\frac{4}{\pi^{1/2}\beta^3} \nonumber \\
&\times& \text{exp}\left(\frac{-r^2 + r^{\prime2}}{\beta^2}\right)i^\ell(-1)^\ell j_\ell(iz),
\label{ell-VanNeck}
\end{eqnarray} 
where $z = 2rr^\prime / \beta^2$ and $j_\ell$ is a spherical Bessel function with a purely imaginary argument. 
The fact that an analytic projection is possible provided the motivation of the choice of Eq.~(\ref{eq:nlocal}).
In arriving at the result of Eq.~(\ref{ell-VanNeck})
use has been made of the relation between the spherical Bessel functions and the Legendre polynomials $P_\ell$:
\begin{equation}
j_\ell(z)=\frac{1}{2i^\ell}\int^{+1}_{-1} dt\ e^{izt}P_\ell(t)  .
\end{equation}

\begin{figure}
\includegraphics[width=3.3in]{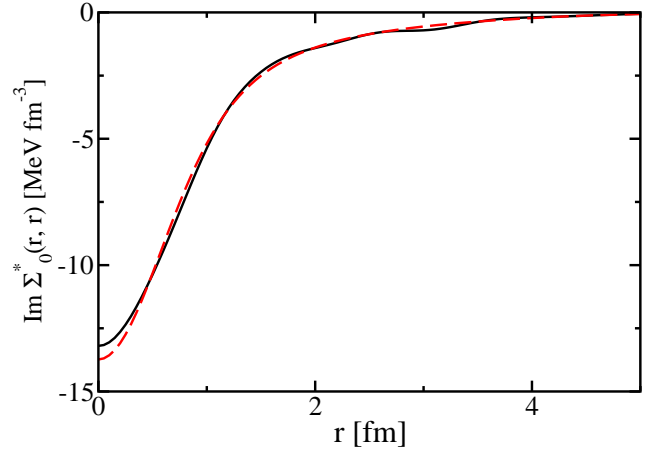}
\caption{(Color online) Diagonal part of CDBonn imaginary self-energy at 65 MeV (solid), and the corresponding parametrized self-energy (dashed). The results shown are for $\ell = 0$.\label{fig:diagx} }
\end{figure}
We have chosen to fit the imaginary part at 65 MeV partly because we expect that only at such energies the imaginary part of the microscopic self-energy has real relevance since the role of LRC is expected to be diminished.
In practise, this means that only the $\ell = 0$ self-energy needs to be represented in terms of Eq.~(\ref{ell-VanNeck}).
If the choice of Eq.~(\ref{eq:nlocal}) is appropriate, the other $\ell$-values will be adequately represented as well.
In Fig.~\ref{fig:diagx} we display the diagonal of the central imaginary part of the self-energy in coordinate space for $\ell=0$ for the CDBonn potential by the solid line.
The fit according to Eq.~(\ref{eq:nlocal}) is quite satisfactory and given by the dashed line.
Quantitative results for diffuseness $a_0$, radius $r_0$, and the non-locality parameter $\beta$ are discussed below.

\begin{figure}
\includegraphics[width=3.2in]{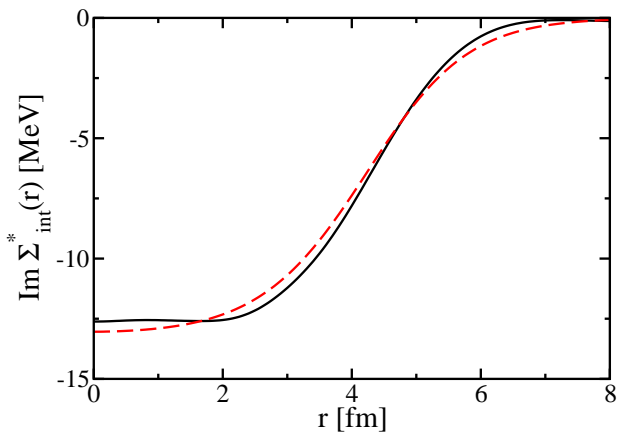}
\caption{(Color online) CDBonn imaginary self-energy at 65 MeV integrated over $r^\prime$ (solid), and the corresponding integrated parametrized self-energy (dashed). The results shown are for $\ell = 0$.
\label{fig:int} }
\end{figure}

Another useful check on the overall relevance of the parametrization of the non-local content of the potential is to integrate over the variable $r'$ in Eq.~(\ref{ell-VanNeck}) to sample the nondiagonal components and compare with the corresponding integral for the CDBonn self-energy.
The result of this procedure is identified by $\Sigma^*_{int}$ and shown in Fig.~\ref{fig:int} as a function of $r$ for the parametrization (dashed) and CDBonn self-energy (solid) for orbital angular momentum $\ell = 0$.
This more stringent test including the sampling of non-diagonal components of the self-energy, still yields a satisfactory representation of the microscopic potential.
It is interesting to note that the shape of the ``local'' potential is more reminiscent of a standard volume absorption.

Another useful quantity to gauge the characteristic of an absorptive potential is the volume integral.
For local potentials this quantity is well-constrained by experimental cross sections~\cite{Charity07,Charity11}.
A recent analysis of the FRPA self-energy reveals useful insights into the comparison of microscopic self-energies with DOM potentials at lower energy~\cite{seth11}.
As in Ref.~\cite{seth11} we define the volume integral for a given orbital angular momentum $\ell$ by
\begin{equation}
J_W^\ell(E) = 4\pi\int{dr\ r^2\int{dr'  r'^{2}\ \text{Im } \Sigma_{\ell}^*(r, r' ; E)}} .
\label{eq:intgs_W} 
\end{equation}
For a local potential it reduces to the standard definition of the volume integral.

\begin{figure}[t]
\includegraphics[width=3.2in]{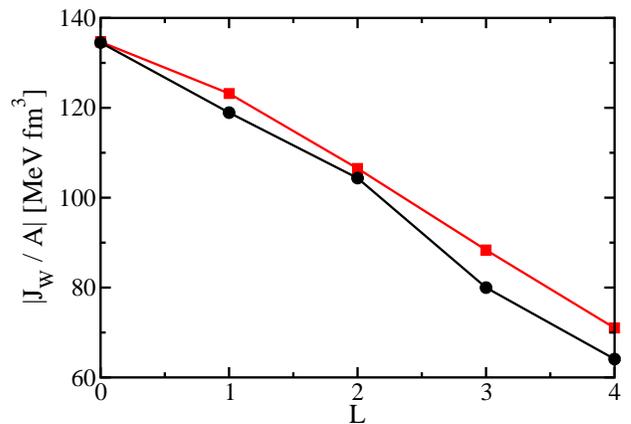}
\caption{ (Color online) Imaginary volume integrals for the CDBonn self-energy at 65 MeV (circles), and the corresponding result for the parametrized self-energy (squares).\label{fig:higherl}}
\end{figure}
The implied $\ell$-dependence of the chosen non-local potential leads to predictions for higher $\ell$-values for this quantity once a fit to the $\ell =0$ component of the self-energy has been made.
The result of the corresponding volume integrals per nucleon are shown in Fig.~\ref{fig:higherl} as a function of the $\ell$-values considered for the CDBonn self-energy.
We employ dots for the CDBonn results and circles for the predictions based on Eq.~(\ref{eq:nlocal}).
The agreement appears very satisfactory and may be useful to extract the properties of the CDBonn self-energy for even higher $\ell$-values without recourse to an explicit calculation.

The properties of the imaginary part of the CDBonn self-energy in terms of its non-locality content are summarized in Table~\ref{tbl:parms} for four different energies, one below and three above the Fermi energy.
In all cases a substantial imaginary part of the CDBonn self-energy is present at the chosen energies.
The parameters are fitted at each energy to reproduce the essential properties of the self-energy including the volume integral for $\ell=0$, as discussed above for the case of 65 MeV.
\begin{table}[b]
\caption{ Parameters from non-local fits to the imaginary part of the proton self-energy at different energies. $W_0$ is in MeV, $r_0, a_0, \beta$ are in fm, and $J_W$ is in units of MeV fm$^3$.\label{tbl:parms}}
\begin{tabular}{ccccccc}
\hline \hline
Energy & $W_0$ & $r_0$ & $a_0$ & $\beta$ & $|J_W / A|$ & $|J_W / A|$ \\
   MeV     &              &            &              &                &                  & CDBonn \\ \hline 
-76 & 36.30 & 0.90 & 0.90 & 1.33 & 193 & 193 \\ 
49 & 6.51 & 1.25 & 0.91 & 1.43 & 73 & 73 \\ 
65 & 13.21 & 1.27 & 0.70 & 1.29 & 135 & 135 \\ 
81 & 23.90 & 1.22 & 0.67 & 1.21 & 215 & 215\\ \hline \hline
\end{tabular}
\end{table}

We observe that the values for the diffuseness generate standard values for the higher energies but are substantially larger for 49 and -76 MeV.
The radius parameter is quite small below the Fermi energy but yields rather standard values at positive energy.
The value of the non-locality parameter is quite a bit larger than typically assumed for real non-local potentials.
Wave function corrections for non-locality in the analysis of $(e,e'p)$ reactions typically assume values of $\beta = 0.85$~fm~\cite{denherder}.
The DOM analysis of Ref.~\cite{Dickhoff10a} introduced a non-local Hartree-Fock potential to allow the calculation of additional properties below the Fermi energy from the spectral functions that are the solutions of the Dyson equation.
The adjusted non-locality parameter in that work corresponded to 0.91 fm.

\begin{figure}
\includegraphics[width=3.2in]{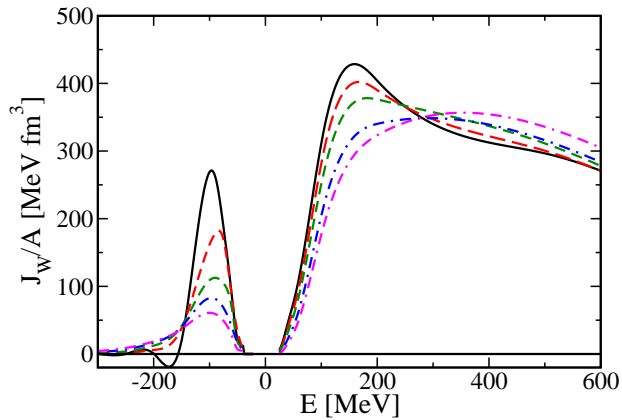}
\caption{ (Color online) Imaginary volume integrals for the CDBonn self-energy as a function of energy for different $\ell$-values: $\ell=0$ (solid), $\ell=1$ (dashed), $\ell=2$ (short-dashed), $\ell=3$ (dash-dot), and $\ell=4$ (dash-dash-dot).\label{fig:Jw}}
\end{figure}

We note that with increasing energy the non-locality parameter decreases suggesting a trend to a more localized potential.
Since for a local potential there is no $\ell$-dependence of the volume integral, we have investigated the behavior of $J_W^\ell$ for different $\ell$-values in a wide energy domain.
The results of this analysis are shown in Fig.~\ref{fig:Jw}.
The degree of non-locality appears to be largest below the Fermi energy with a substantial separation between the different $\ell$-values. 
The result for $\ell=0$ also demonstrates that it is possible to have the ``wrong'' sign for the volume integral.
This can happen because the microscopic self-energy develops negative lobes off the diagonal and a positive volume integral cannot be guaranteed as a result, as must be the case for a local potential.
Although the imaginary part above the Fermi energy is negative, it is conventional to plot the imaginary volume integral as a positive function of energy~\cite{Charity11}.
At positive energy the volume integrals for different $\ell$ at first exhibit a spread although not as large as below the Fermi energy.
Above 300 MeV however, the curves apparently become similar suggesting a trend to a more local self-energy.

\section{Conclusions}
\label{sec:conc}

The properties of the microscopic self-energy of nucleons derived from the realistic CDBonn interaction have been studied for ${}^{40}$Ca.
The calculation involves a two-step procedure starting with the calculation of a $\mathcal{G}_{NM}$-matrix interaction in nuclear matter for a fixed energy and density.
In a second step, the Fermi structure of the finite nucleus is incorporated by expanding the finite-nucleus $\mathcal{G}_{FN}$-matrix in the nuclear matter one, including up to second-order terms.
The self-energy is obtained by including the corresponding self-energy terms with imaginary parts above and below the Fermi energy, with associated real parts obtained from the appropriate dispersion relations.
The analysis of the solutions of the Dyson equation below the Fermi energy includes spectral functions calculated in momentum space, momentum distributions, quasihole properties (including spectroscopic factors), natural-orbit properties, the nuclear charge density, and the energy of the ground state of ${}^{40}$Ca.

An important motivation for the present work is to generate insight from microscopic calculations what functional forms of the nucleon self-energy can be employed fruitfully in the analysis of experimental data in the DOM framework.
Recent DOM work has also focused on ${}^{40}$Ca and a parallel comparison with DOM results is therefore pursued throughout the paper.

Nucleon spectral functions for the CDBonn potential exhibit similar features as those from earlier work for ${}^{16}$O using the Bonn-B potential although the former interaction appears somewhat softer.
This leads to a less pronounced presence of high-momentum components at very negative energies.
The energy distribution of these momenta is somewhat different than the one generated by the DOM self-energy although the fraction of high-momentum particles is about 10\% in both calculations.
 
Since noninteracting intermediate states are employed in the CDBonn self-energy and therefore LRC are not well incorporated, there is no imaginary part in a substantial region around the Fermi energy.
As a result, only the lowest $s_{1/2}$ state is broadened in accordance with experiment, whereas all other quasihole states are represented by discrete states. 
The DOM calculation exhibits a more realistic distribution of the sp strength including appropriate widths for $p$-states as well.
The location of the quasihole states in the CDBonn calculation is in reasonable agreement with experiment but the particle-hole gap is larger than experiment.
The associated spectroscopic factors are close to 0.9 consistent with the 10\% fraction of high-momentum nucleons.
The DOM spectroscopic factors are about 0.2 smaller since the DOM self-energy includes a strong coupling to the nuclear surface leading to better agreement with the analysis of $(e,e'p)$ reactions.

The calculation of natural orbits demonstrates that the largest occupation numbers are close to 0.9 very similar to a recent DOM calculation even though substantial differences in spectroscopic factors occur, as discussed above.
It appears that nuclear natural orbits always generate such occupation numbers in contrast with finite drops of ${}^3$He atoms, where they can be substantially smaller in accordance with the much stronger repulsion of the underlying interaction.

The nuclear charge density from the CDBonn self-energy exhibits too small a radius and too much charge at the origin but is otherwise not too dissimilar from the DOM results.
An important difference however, is the presence of a substantial non-local imaginary self-energy below the Fermi energy in the microscopic calculations.
This leads to a good convergence with orbital angular momentum for the number of particles which amounts to 19.3 neutrons when $\ell_{max}=4$.
No such convergence is obtained with DOM calculations on account of the locality of the imaginary self-energy, thereby overestimating the number of particles.
We therefore conclude that the introduction of non-locality in the imaginary DOM potentials in the future is an essential ingredient.

The distribution of high-momentum nucleons from the CDBonn calculation leads to their large contribution of 67\% to the energy per particle in agreement with earlier observations for ${}^{16}$O.
The more realistic distribution of high-momenta leads to about 2 MeV more binding per nucleon than from the DOM self-energy but still 2 MeV above the experimental result pointing to the need of an improved treatment of intermediate states in the self-energy and the consideration of higher-order contributions in the nuclear-matter $\mathcal{G}_{NM}$-matrix interaction.

We also performed an investigation of the CDBonn self-energy at positive energy.
The lack of higher partial waves considered in the calculation leads to an underestimate of the total cross section.
A comparison with the same number of partial waves in the DOM calculation demonstrates that above 70 MeV there is reasonable agreement between the two approaches suggesting that volume absorption is well represented by the present calculation but surface contributions (LRC) are missing.
The differential cross section are not in good agreement with the corresponding DOM results (which generate good fits to the data when all contributing partial waves are included).
This lack of agreement is worrisome but  suggest that a lot more work is needed, in particular the inclusion of LRC.

Finally, an analysis of the non-locality of the imaginary part to the CDBonn self-energy reveals that its main properties can be quite well represented by a gaussian non-locality.
Typical non-locality parameters are somewhat larger than those found in the literature.
Volume integrals indicate that non-locality is very important below the Fermi energy.
Above the Fermi energy, it is initially substantial but appears to weaken at higher energies.

\acknowledgments This work was partly supported by the U.S.
National Science Foundation under grant PHY-0968941, grant No. FIS2008-01661 from MEC and FEDER (Spain), and grant 2009SGR-1289 from Generalitat de Catalunya.

\bibliography{level_de}


\end{document}